\documentclass[aps,pre,showpacs,twocolumn,numerical]{revtex4-1}

\usepackage[colorlinks=true,urlcolor=blue,citecolor=red,linkcolor=red,bookmarks=true]{hyperref}

\usepackage[utf8]{inputenc}
\usepackage[T1]{fontenc}
\usepackage[normalem]{ulem}
\usepackage{epstopdf, epsfig}

\usepackage{amsmath,amssymb,amsfonts}
\usepackage{graphicx}
\usepackage{xspace}
\usepackage{textcomp}
\usepackage{color}
\usepackage{tikz}
\usetikzlibrary{patterns}
\usetikzlibrary{calc}

\tikzset{
    ncbar angle/.initial=90,
    ncbar/.style={
        to path=(\tikztostart)
        -- ($(\tikztostart)!#1!\pgfkeysvalueof{/tikz/ncbar angle}:(\tikztotarget)$)
        -- ($(\tikztotarget)!($(\tikztostart)!#1!\pgfkeysvalueof{/tikz/ncbar angle}:(\tikztotarget)$)!\pgfkeysvalueof{/tikz/ncbar angle}:(\tikztostart)$)
        -- (\tikztotarget)
    },
    ncbar/.default=0.5cm,
}

\tikzset{square left brace/.style={ncbar=0.5cm}}
\tikzset{square right brace/.style={ncbar=-0.5cm}}

\tikzset{round left paren/.style={ncbar=0.5cm,out=120,in=-120}}
\tikzset{round right paren/.style={ncbar=0.5cm,out=60,in=-60}}

\newif\ifhyper
\hypertrue
\ifhyper
\hypersetup{
  citecolor = {green},
  urlcolor = {blue} 
} 

\newcommand{\p}{\partial}

\newcommand{\vx}{\vec{x}}

\newcommand{\vrr}{\vec{r}}
\newcommand{\vv}{\vec{v}}
\newcommand{\vnabla}{\vec{\nabla}}

\newcommand{\vu}{\vec{u}}

\newcommand{\vf}{\vec{f}\,}
\newcommand{\vJ}{\vec{J}}

\newcommand{\hp}{{\hat p}}

\newcommand{\hy}{{\hat y}}

\newcommand{\bq}{{\bf q}}
\newcommand{\bp}{{\bf p}}

\newcommand{\bk}{{\bf k}}
\newcommand{\bx}{{\bf x}}
\newcommand{\by}{{\bf y}}
\newcommand{\bz}{{\bf z}}
\newcommand{\vq}{{\vec{q}}}
\newcommand{\vp}{{\vec{p}}}
\newcommand{\vk}{{\vec{k}}}

\newcommand{\sref}[1]{Sec.~\ref{#1}}

\newcommand{\aref}[1]{Appendix~\ref{#1}}

\newcommand{\Eq}[1]{Eq.~(\ref{#1})}
\newcommand{\eq}[1]{(\ref{#1})}
\newcommand{\Eqs}[1]{Eqs.~(\ref{#1})}

\newcommand{\ie}{{\it i.e.}\xspace}
\newcommand{\eg}{{\it e.g.}\xspace}

\usepackage{abbrevs}
\usepackage{etoolbox}
\newabbrev\RG{Renormalisation Group (RG)}[RG]
\newabbrev\NPRG{Non-Perturbative (also named functional) Renormalisation Group (NPRG)}[NPRG]
\newabbrev\NS{Navier-Stokes (NS)}[NS]
\newabbrev\PI{one particle-irreducible (1-PI)}[1-PI]
\newabbrev\rhs{right-hand side (r.h.s.)}[r.h.s.]
\newabbrev\lhs{left-hand side (l.h.s.)}[l.h.s.]

\def\mean#1{\left< #1 \right>}
\def\abs#1{\left| #1 \right|}

\robustify{\RG}
\robustify{\NPRG}
\robustify{\NS}
\robustify{\PI}
\robustify{\rhs}
\robustify{\lhs}

\makeatletter
\renewcommand\maybe@space@{%
  \maybe@ictrue 
  \expandafter   \@tfor
    \expandafter \reserved@a
    \expandafter :%
    \expandafter =%
                 \nospacelist
                 \do \t@st@ic
  \ifmaybe@ic 
    \space
  \fi
}
\makeatother

\begin{document}

\title{Breaking of scale invariance in the time dependence of correlation functions in isotropic and homogeneous turbulence}

\author{Malo Tarpin}
\affiliation{Universit\'e Grenoble Alpes and CNRS, LPMMC,  38000 Grenoble, France }
\author{L\'eonie Canet}
\affiliation{Universit\'e Grenoble Alpes and CNRS, LPMMC,  38000 Grenoble, France }

\author{Nicol\'as Wschebor}
\affiliation{Instituto de F\'isica, Facultad de Ingenier\'ia, Universidad de la Rep\'ublica, J.H.y Reissig 565, 11000 Montevideo, Uruguay}

\begin{abstract}
In this paper, we present  theoretical results on the statistical properties of stationary,
 homogeneous and isotropic turbulence   in incompressible flows in three dimensions. Within the 
 framework of the Non-Perturbative Renormalization Group, we derive a closed renormalization flow equation 
 for a generic  $n$-point correlation (and response) function  for large wave-numbers with respect to the 
 inverse integral scale. The  closure is obtained from a controlled expansion and relies on extended symmetries 
 of the Navier-Stokes field theory. It  yields the exact leading behavior of the flow equation at large wave-numbers
  $|\vp_i|$, and for arbitrary time differences $t_i$  in the stationary state. Furthermore, we obtain the form  of 
  the general solution of the corresponding fixed point equation, which yields the analytical form of the leading
   wave-number and time dependence of $n$-point correlation functions,  for large wave-numbers and both for small
     $t_i$ and in the limit $t_i\to \infty$. At  small $t_i$, the leading contribution  at large wave-number is 
     logarithmically equivalent to $-\alpha (\varepsilon L)^{2/3}|\sum t_i \vp_i|^2$, where $\alpha$ is a non 
     universal constant, $L$ the integral scale and $\varepsilon$ the mean energy injection rate. For the 2-point 
     function, the $(t p)^2$ dependence is known to  originate from the  sweeping effect.  The derived formula 
     embodies the generalization of the effect of sweeping to $n-$point   correlation functions.  At large wave-number
      and large $t_i$, we show that  the $t_i^2$ dependence in the leading order contribution crosses over to a $|t_i|$
       dependence.  The expression of the correlation functions in this regime was not derived before, even for the 2-point 
       function.  Both predictions can be tested in direct numerical simulations and in experiments.
\end{abstract}

\maketitle

\section{Introduction}

Correlations of the velocity are central objects in the statistical theory of turbulence. 
 We focus in this paper on isotropic and homogeneous fully developed turbulence in three-dimensional incompressible flows.
 A common measure of spatial correlations is provided by the structure functions $S_n(r)$,
  which are the moments of order $n$ of equal-time longitudinal velocity increments
 $\delta u(r) = (\vv(t,\vx) - \vv(t,\vx+\vrr))\cdot\hat{r}$. The structure functions exhibit 
 universal power-law behaviors  $S_n(r)\sim r^{\xi_n}$ in the inertial range of scales, that is in between the integral scale $L$ where
 energy is injected and the Kolmogorov scale $\eta$ where it is dissipated by molecular friction. 
  Whereas Komogorov theory (K41)~\citep{Kolmogorov41}
 predicts linearly increasing exponents $\xi_n = n/3$, experiments and direct numerical simulations
 early evidenced systematic deviations to K41 scalings, which are hallmarks of multi-scaling, and intermittency. 
   The turbulent steady state appears as a critical non-equilibrium (driven-dissipative) state, 
    characterized by non-standard scale invariance, which 
 renders its study particularly challenging.
  
  Beyond equal-time statistics, the time dependence of correlation functions is also 
 of primary interest. The typical decorrelation time of the velocity,
  and also the behavior of the frequency energy
 spectrum, have early been debated.
  Indeed, the results derived by direct application of Kolmogorov original local similarity hypothesis are in contradiction with 
 results stemming from other theoretical arguments taking into account the  random sweeping of small eddies by larger ones~\citep{Heisenberg48, Kraichnan59,Tennekes75}.
 In particular, whereas the local similarity hypothesis predicts an energy spectrum behaving as $\omega^{-2}$
 for Eulerian velocities, the sweeping effect leads instead to the power law $\omega^{-5/3}$.
From the analysis of simple models of advection, Kraichnan deduced that the sweeping effect should yield
 for the two-point correlation function a Gaussian in the variable $tp$ where $p$ is the wave-number and $t$ the time delay.
This behavior has later been  observed in many numerical simulations~\citep{Orszag72, Sanada92, He04, Favier10, Canet17}
  and also in experiments~\citep{Poulain06}.
 Kraichnan extended this analysis to the three-point correlations in some specific time and wave-vector configurations,
  but the general expression of the sweeping effect for $n$-point correlation functions is not known. 
 The Gaussian in  $tp$ for the two-point function was also confirmed in a \RG study of \NS equation with an effective viscosity \cite{Antonov94}.
 
 Because it is a source of breaking of scale-invariance, the sweeping effect has hindered progress in application of \RG to turbulence. 
  This effect, already recognized by Kraichnan \cite{Kraichnan64}, was circumvented using quasi-Lagrangian velocities in \cite{Belinicher87}
   and an extended form of the Galilean symmetry  in \cite{Adzhemyan94}.
  These lines of work allowed both groups to obtain results on the equal-time statistics of turbulence,
   in particular that there cannot be any strictly perturbative correction
  to K41 scaling \cite{Lvov95a}.
 
 Recently, turbulence has been re-visited using an alternative formulation of the \RG, based on \NPRG 
 techniques~\citep{Canet15, Canet16, Canet17}.  In this approach, a closed \RG flow equation for the two-point correlation function, which is asymptotically exact in the limit of large wave numbers, was obtained.
 At variance with standard closure schemes, such as DIA (Direct-Interaction Approximation) \cite{Kraichnan59,Kraichnan64}, or EDQNM (Eddy-Damped Quasi-Normal Markovian approximation) \cite{Orszag70}, 
 this approach does not involve any  uncontrolled truncation. It consists in an expansion of the flow equation at large wave-numbers, whose leading term can be computed exactly. The universal space- and time-dependent two-point correlation function,
  obtained as the solution of the fixed-point equation,
  confirmed the Gaussian in the variable  $tp$ for small time delays, in accordance with previous findings. But in contrast 
  with former approaches, in the \NPRG framework, this result 
  was derived  from the full \NS equation in a rigorous way, as the exact leading contribution at large wave-numbers and small times.
  
 In the present paper, we use the \NPRG framework to derive the generic space and time dependence of 
 any $n-$point generalized  (velocity and response) correlation function in a stationary homogeneous and isotropic
  turbulent state in three dimensions. We obtain an analytical expression which is exact in the limit of large wave-numbers 
  (and for  non-exceptional wave-vector configurations), in both regimes of small and large time delays. Such rigorous  theoretical results are scarce in the context of turbulence. The expression obtained for generalized $n$-point correlation function
 $G^{(n)}$
  explicitly breaks standard scale invariance.
  The physical origin of this breaking is identified as the sweeping effect at small time delays, but takes a different form at large time delays, suggesting  a different mechanism at play. This calculation is performed in the Eulerian framework. 
  The leading term in wave-numbers of $\log G^{(n)}$, which is of order $p^2$, is computed exactly.
 The scaling of $G^{(n)}$ (the power-law part), corresponds to a sub-leading term, of order $\log p$, in this expansion, 
  and it  is not captured exactly in the present leading order calculation. 
  As a consequence, this calculation is not appropriate to determine intermittency
   corrections to the scaling exponents $\xi_n$. This requires the derivation of the sub-leading term in the expansion in wave-numbers, which is left for future investigation.

The remainder of the paper is organized as follows. The field theory and the \NPRG formalism associated with the forced \NS equation are reviewed in \sref{secNS}. The symmetries of this field theory,  in particular  their extended forms, are recalled in \sref{secSym}, with the corresponding Ward identities.
 In \sref{secFlow}, we derive the first main result of the paper, which is the closed flow equation for 
 any generalized correlation functions, which is the exact leading order contribution at large wave-numbers for non-exceptional wave-vector configurations.
In section \sref{sec2point}, we study the solution of this equation for the 2-point correlation function. We give its
 behavior for small time delays, which was already obtained in \citep{Canet17}, and further derive its behavior
  at large time delays. In \sref{secSol}, we derive the second main result of the paper,
   which is the analytical form of the solution of the fixed point equation for any correlation function, 
   yielding the form of its space and time dependence,  in both the regimes of small and of large time delays. We elaborate on possible tests of these predictions. Several technicalities and calculations are detailed in the appendices.

\section{Navier-Stokes field theory and NPRG formalism}
\label{secNS}

Our starting point is the \NS equation in the presence of an external stirring force whose role is to maintain a 
 turbulent steady state:
\begin{equation}
  \p_t \vv+ \vv \cdot \vnabla \vv=-\frac 1\rho 
\vnabla p +\nu \nabla^2 \vv +\vf \, ,
\label{eq:ns}
\end{equation}
 where $\nu$ is the kinematic viscosity and 
$\rho$ the density of the fluid.  
 The velocity field $\vv$, the pressure field $p$, and the  
force $\vec f$ depend on the 
space-time coordinates $\bx \equiv (t,\vx)$.  
 We focus on incompressible flows, satisfying 
\begin{equation}
 \vec\nabla \cdot \vv = 0 \, .
\label{eq:compress}
\end{equation}
We are interested in universal properties, which are the  properties characterizing all turbulent steady-states, that do not depend on the precise details of the  fluid or of the forcing mechanism. 
 To study universal properties, one can conveniently  consider a stochastic forcing
 and average over its realisations. This forcing is chosen with a Gaussian distribution of
 zero mean and variance
\begin{equation}
 \langle f_\alpha(t,\vx)f_\beta(t',\vx\,')\rangle=2 \delta_{\alpha\beta}\delta(t-t')N_{L^{\text{-}1}}(|\vx-\vx\,'|) \, ,
\end{equation}
where $\langle \rangle$ denotes the ensemble average. This correlator is local in time, to preserve Galilean invariance,
 and it is  concentrated, in Fourier space, on the inverse of the integral scale $L$.
  Let us emphasize that
  at variance with many perturbative \RG studies, one does not need to impose  a power-law forcing 
 whose role is to introduce a formal expansion parameter $\epsilon$. 
   The profile $N_{L^{\text{-}1}}$ is a smooth regular function (both in the IR and in the UV),
  representative of the physical large-scale forcing.
  It can be chosen diagonal in component space,
 without loss of generality,  because of incompressibility.
 
The stochastic \NS equation (\ref{eq:ns}) can be represented as a field theory following the
 standard Martin-Siggia-Rose-Janssen-de Dominicis (MSRJD) response functional formalism \citep{Martin73,Janssen76,Dominicis76}, which yields the generating functional \citep{Canet16}
\begin{align}
 {\cal Z}[\vJ,\bar{\vJ},K,\bar{K}] =  &\int \mathcal{D}[\vv, \,p, \,\bar{\vv}, \,\bar p]\,
 \, e^{-{\cal S}_0[\vv,\bar{\vv},p,\bar p] -\Delta {\cal  S}_{0}[\vv,\bar{\vv}] } \nonumber\\
 \times &e^{\int_{\bx}\{ \vJ\cdot \vv+\bar{\vJ}\cdot \bar{\vv}+K p+\bar K\bar p\}} 
 \, ,\label{eq:Z}
\end{align}
with notation $\int_\bx \equiv \int d^d \vx dt$, and 
where $\vJ, K, \bar \vJ, \bar K$ are the sources for the velocity, pressure and the associated response fields.
 For  completeness, a presentation of the MSRJD response formalism and the mapping of the NS equation to
  a field theory is detailed in Appendix~\ref{AP:MSRJD}. Let us note that this standard derivation  assumes 
  existence and uniqueness of weak solutions of the three-dimensional NS equations, which is actually
   questioned mathematically in some cases, see \eg \cite{Buckmaster17} and related comments in Appendix~\ref{AP:MSRJD}.
 Generalized correlation  functions are obtained by taking functional derivatives of ${\cal Z}$ with respect to the sources.
The \NS action, separated in two parts for later purpose, is given by
\begin{align}
 {\cal S}_0[\vv,\bar{\vv},p,\bar p] &= \int_{\bx}\Big\{\bar p (\bx)\,\p_\alpha v_\alpha(\bx) + \bar v_\alpha(\bx)\Big[ \p_t 
v_\alpha(\bx) \nonumber\\
 &\quad-\nu \nabla^2 v_\alpha(\bx) +v_\beta (\bx)\p_\beta v_\alpha(\bx)+\frac 1\rho \p_\alpha p (\bx) \Big]\Big\}\nonumber\\
 \Delta {\cal S}_{0}[\vv,\bar{\vv}] &=-\int_{t,\vec x,\vec x'}\bar 
v_\alpha(t,\vec x) N_{L^{\text{-}1}}(|\vec x-\vec x'|)\bar v_\alpha(t,\vec  x')\, .
\label{eq:NSaction}
\end{align}

This field theory can be studied using \NPRG techniques. The \NPRG is a modern implementation of
 Wilson's original idea of the \RG \citep{Wilson74}, conceived to efficiently average over fluctuations, even when they
 develop at all scales, as in standard critical phenomena \citep{Berges02,Kopietz10,Delamotte12}. It is a powerful 
 method to compute the properties of strongly
  correlated systems,  which can reach high precision levels \citep{Canet03b,Benitez12},
  and can yield fully non-perturbative results, at equilibrium \citep{Grater95,Tissier06,Essafi11} and also 
  for non-equilibrium systems \citep{Canet04a,Canet05,Canet10,Canet11a,Berges12}, restricting to a few classical statistical
   physics applications. The \NS field theory was studied
 using \NPRG methods in \citep{Tomassini97,Monasterio12,Canet16}. We here exploit the formalism
    constructed and expounded in details in \citep{Canet16}.
 In summary, the selection of fluctuation modes is achieved by replacing the quadratic term $\Delta {\cal S}_0$ in (\ref{eq:NSaction})    by a regulator term $\Delta {\cal S}_\kappa$, which depends
 on a running wave-number scale $\kappa$. This regulator term consists of two parts, which can be interpreted respectively
  as an effective forcing and an effective viscosity:
\begin{align}
 \Delta {\cal S}_\kappa[\vv,\bar \vv] =&-\int_{t,\vec x,\vec x'}\bar 
v_\alpha(t,\vec x) N_{\kappa}(|\vec x-\vec x'|)\bar v_\alpha(t,\vec 
x') \nonumber\\
&+\int_{t,\vec x,\vec x'}\bar v_\alpha(t,\vec x) R_{\kappa}(|\vec 
x-\vec x'|) v_\alpha(t,\vec x') \, .
\label{eq:deltaSk}
\end{align}
The most important feature of $N_\kappa$ and $R_\kappa$ that will be exploited in the following is that they 
  rapidly vanish (typically exponentially) for  wave-numbers large compared to the \RG scale $\kappa$ 
 \begin{equation}
   R_\kappa(\vq) \xrightarrow[|\vq|\gg \kappa]{} 0\, , \quad \quad N_\kappa(\vq) \xrightarrow[|\vq|\gg \kappa]{} 0 \,. \label{eq:limitreg}
 \end{equation}
Another important property is that they are required to regularize the flow at small wave-numbers (see \citep{Berges02,Canet16}).

In the presence of the regulator $\Delta {\cal S}_\kappa$, the generating functional defined in (\ref{eq:Z}) becomes scale dependent, 
  and is denoted ${\cal Z}_\kappa$. 
 The average of the velocity (and response velocity) fields
 can be obtained through derivatives of ${\cal W}_\kappa = \ln {\cal Z}_\kappa$, as
\begin{equation}
  u_\alpha({\bx}) = \langle v_\alpha({\bf x}) \rangle = \frac{\delta {\cal 
W}_{\kappa}}{\delta J_\alpha({\bf x})}  \, \, , \, \,
  \bar u_\alpha({\bf x}) = \langle \bar v_\alpha({\bf x}) \rangle = \frac{\delta 
{\cal W}_{\kappa}}{\delta \bar J_\alpha({\bf x})}\, , \nonumber
\end{equation} 
(and similarly for the pressure fields). ${\cal W}_\kappa$ is the generating functional of connected 
correlation functions, which are the generalizations of cumulants for a field theory.
The average effective action $\Gamma_\kappa$ 
(which is the generating functional of \PI correlation functions, see~\cite{Amit84})
 is defined as the Legendre transform
 of ${\cal W}_\kappa$, up to the regulator term:
\begin{align}
\Gamma_\kappa[\vu,\bar\vu,p,\bar p] &+ {\cal W}_\kappa[\vJ,\bar{\vJ},K,\bar K] = 
\int_{\bx} \! \Big\{ \vJ\cdot \vv+\bar{\vJ}\cdot \bar{\vv} \nonumber\\
+ & K p+\bar K\bar p\Big\} -\int_{t,\vx,\vx'} \Big\{ \bar u_\alpha \,R_{\kappa} \,u_\alpha 
- \bar u_\alpha \,N_{\kappa} \,\bar u_\alpha \Big\} \, .
\label{eq:legendre}
\end{align}
The flow of  $\Gamma_\kappa$, which means the evolution of the latter with the \RG scale $\kappa$, is given by the Wetterich equation \citep{Wetterich93}
\begin{equation}
\p_\kappa \Gamma_\kappa = \frac{1}{2}\, \int_{\bx,\by}\! 
 \p_\kappa [{\cal R}_\kappa]_{ij}(|\bx-\by|)   \Big[\Gamma_\kappa^{(2)} + {\cal R}_\kappa\Big]_{ji}^{-1}\, ,
\label{eq:dkgam}
\end{equation}
 where  $\Gamma_\kappa^{(2)}$ is the Hessian of $\Gamma_\kappa$ and the regulator matrix  $[{\cal R}_\kappa]$ is defined as
\begin{equation}
 [{\cal R}_\kappa]_{ij} = \frac{\delta^2 \Delta {\cal S}_\kappa}{\delta \varphi_i \delta \varphi_j} \, ,
\label{eq:calR}
\end{equation}
with $\varphi_i$ standing for any of the average velocity and pressure fields $u_\alpha$, $\bar{u}_\alpha$, $p$ or $\bar p$.
The \RG flow equation \eq{eq:dkgam} is exact. Its initial condition corresponds to the `microscopic' model, 
 which is the \NS equation. The flow is hence initiated at a very large wave-number $\Lambda$ at which 
 the continuous description of the fluid dynamics in terms of \NS equation starts to be valid. At this scale, one can show
  that $\Gamma_\Lambda$ identifies with the bare action $\Gamma_\Lambda={\cal S}$, since no fluctuation is yet incorporated.
   When $\kappa\to 0$, the regulator is removed and one obtains the actual properties of the model, when all fluctuations
   have been integrated over.  \Eq{eq:dkgam} provides the exact interpolation between these two scales. 
In general, the functional partial differential equation  (\ref{eq:dkgam}) cannot be solved exactly, and its study requires  some approximations.

   In this work, we do not integrate the flow equation (\ref{eq:dkgam}) from given initial conditions, but rather
  search for fixed-point solutions.
  Indeed, the turbulent steady-state is known to exhibit some form of scale invariance in the inertial range.
  For standard critical phenomena, scale invariance corresponds to fixed-points of the \RG flow equations (which is explained in more details in Appendix  \ref{AP:RG} for unfamiliar readers).  Hence, we will restrain ourselves to this particular class 
   of solutions of  (\ref{eq:dkgam}). Fixed-point solutions depend on $\kappa$ only through the scaling of space-time and of the
   fields. For usual critical phenomena, this choice leads to scale invariance of $\Gamma_\kappa$ for wave-numbers much larger than $\kappa$.
   This is not the case for turbulence. Indeed, the main result of the paper is
   the exact equation which replaces scale invariance for turbulence. 
   It is important to stress that, similarly to standard critical phenomena where there are finite-size corrections to scale invariance,
   there are also finite-size corrections to this equation in turbulence, and hence to the leading behavior at large wave-numbers of the correlations functions. In particular, these corrections, not captured
   by the present analysis,  determine the intermittency effects {\it at equal time}, \ie for the structure functions.

 The regulator matrix \eq{eq:calR} has only two non-vanishing elements (see \Eq{eq:deltaSk}), which are in the velocity sector. 
 As a consequence, one can show that the pressure sector is decoupled, and furthermore that it is not renormalized \citep{Canet16}. The flow equation \eq{eq:dkgam} effectively reduces to the space of velocity and response velocity fields
 only: that is $i, j =1, 2$ with $\varphi_1 = u_\alpha$, $\varphi_2 = \bar u_\alpha$. 

 The results we present in the following are more conveniently expressed in terms of the connected correlation functions (as was the case for the $2-$point function in \citep{Canet17}). Hence, our starting point is the exact flow equation for ${\cal W}_\kappa$, (which is similar to the Polchinski equation \citep{Polchinski84}):
\begin{align}
 \p_\kappa {\cal W}_\kappa = &- \frac{1}{2}\, \int_{\bx,\by}\! \p_\kappa [{\cal R}_\kappa]_{ij}({\bx-\by}) \,\Big\{\frac{\delta^2 {\cal W}_\kappa}{\delta j_{i}({\bx}) \delta j_{j}({\by})} \nonumber\\
 &+ \frac{\delta {\cal W}_\kappa}{\delta j_{i}({\bx})} \frac{\delta {\cal W}_\kappa}{\delta j_{j}({\by})}\Big\}\label{eq:dkw} \, ,
\end{align}
where as previously $i,j=1,2$ with $j_1=J_\alpha$ and $j_2=\bar J_\alpha$.
${\cal W}_\kappa$ and $\Gamma_\kappa$ are the generating functionals  of respectively the connected and the \PI correlation functions.
 This means that  any
  connected (respectively \PI) generalized correlation functions  can be obtained by taking functional derivatives of ${\cal W}_\kappa$ (respectively $\Gamma_\kappa$).
  As a consequence, one can deduce the flow equation for a generic connected $n$-point correlation and response function by taking the $n$ corresponding functional derivatives of \Eq{eq:dkw}. 
  This yields an equation which is exact, but which involves $(n+1)$ and $(n+2)$ functions, such that one has to solve an infinite hierarchy of flow equations. 
 In most applications, this hierarchy is closed by simply truncating higher-order vertices, or proposing an ansatz for $\Gamma_\kappa$ \cite{Berges02}.
 We show in the following that for \NS, the flow equation for a generic $n$-point function can be closed using 
  a controlled expansion in wave-numbers, without resorting to uncontrolled truncations.

 We finish this section by introducing some useful notation for the $n$-point functions.
 The $n$-point vertex (\PI) functions are  defined as
\begin{equation}
 \Gamma_{\alpha_1\dots\alpha_{n}}^{(n)}[{\bx_1}, \dots, {\bx_n};\varphi] =  \frac{\delta^{n} \Gamma_\kappa}{\delta \varphi_{\alpha_1}({\bx_1}) \dots \delta \varphi_{\alpha_{n}}({\bx_{n}})} \, ,
\end{equation}
where  $\alpha_k$ stands for both the field index (1 or 2) and the space component. 
 They are the renormalized interactions of the theory.
 Notice that in this definition, $\Gamma_{\alpha_1\dots\alpha_{n}}^{(n)}$ is still a functional of the fields, which is materialized by the square brackets and the explicit $\varphi$ dependency.
 Another useful notation  is $\Gamma^{(m,n)}_{\alpha_1\dots\alpha_{n+m}}[{\bx_1}, \dots, {\bx_{n+m}};\varphi]$ where the $m$ first derivatives are with respect
 to $u_{\alpha_k}$ and the $n$ last with respect to  $\bar u_{\alpha_k}$ ($\alpha_k$ being only the space component in this notation).
 We indicate that a vertex function is evaluated at zero fields  using the notation
\begin{equation}
 \Gamma_{\alpha_1\dots\alpha_{n}}^{(n)}({\bx_1}, \dots, {\bx_n})\equiv\Gamma_{\alpha_1\dots\alpha_{n}}^{(n)}[{\bx_1}, \dots, {\bx_n};\varphi=0]
\end{equation}
(and accordingly for $\Gamma^{(m,n)}$). 
Furthermore, we denote by 
 $\tilde \Gamma^{(n)}_{\alpha_1\dots\alpha_{n}}({\bp_1}, \dots, {\bp_{n}})$
 the Fourier transform of $\Gamma_{\alpha_1\dots\alpha_{n}}^{(n)}({\bx_1}, \dots, {\bx_n})$, with $\bp \equiv (\omega,\vp)$.
Because of translational invariance in space and time, this Fourier transform takes the form
\begin{align}
 \tilde \Gamma^{(n)}_{\alpha_1\dots\alpha_{n}}({\bp_1}, \dots, {\bp_{n}}) &= (2\pi)^{d+1}\delta\left(\sum \omega_i\right)\delta^d\left(\sum \vp_i\right) \nonumber\\
 &\times \bar\Gamma^{(n)}_{\alpha_1\dots\alpha_{n}}({\bp_1}, \dots, {\bp_{n-1}}) \, ,
\end{align}
 that is, the total momentum and frequency are conserved.

Finally,  similar conventions are used for the generalized connected correlation functions
\begin{equation}
 G_{\alpha_1\dots\alpha_{n}}^{(n)}[{\bx_1}, \dots, {\bx_n};j]=  \frac{\delta^{n} {\cal W}_\kappa}{\delta j_{\alpha_1}({\bx_1}) \dots \delta j_{\alpha_{n}}({\bx_{n}})} \, ,
\end{equation}
or alternatively $G^{(m,n)}_{\alpha_1\dots\alpha_{n+m}}[{\bx_1}, \dots, {\bx_{n+m}};j]$. 
Accordingly, we define
$ G_{\alpha_1\dots\alpha_{n}}^{(n)}({\bx_1}, \dots, {\bx_n})$ and $ G_{\alpha_1\dots\alpha_{n}}^{(m,n)}({\bx_1}, \dots, {\bx_n})$ as the previous correlation functions evaluated at zero field,
 and their Fourier transforms as
\begin{align}
 \tilde G^{(n)}_{\alpha_1\dots\alpha_{n}}({\bp_1}, \dots, {\bp_{n}}) &= (2\pi)^{d+1}\delta\left(\sum \omega_i\right)\delta^d\left(\sum \vp_i\right) \nonumber\\
 &\times \bar G^{(n)}_{\alpha_1\dots\alpha_{n}}({\bp_1}, \dots, {\bp_{n-1}}) \, .
\end{align}
 The set of connected and \PI correlation functions are inter-related. Indeed, any $n$-point connected correlation function
 $G^{(n)}$ can be constructed as a sum of tree diagrams whose vertices are the \PI functions $\Gamma^{(k)}$, $2<k\leq n$ 
 and the edges the propagators $G^{(2)}$.

\section{Symmetries and Ward identities}
\label{secSym}

The symmetries of the \NS action \eq{eq:NSaction} and the associated Ward identities are studied in details in \citep{Canet15,Canet16}.
Besides the global symmetries, rotational, translational and Galilean invariance, the \NS action also possesses
 a local-in-time, or time-gauged, form of Galilean symmetry. This time-gauged Galilean symmetry, and the related Ward identities, are well-known in the context of field theoretical
   studies of turbulence \citep{Adzhemyan94,Adzhemyan99,Antonov96,Berera07}.
 It corresponds to the following field transformation 
\begin{align} 
 \delta v_\alpha(\bx) =-\dot{\epsilon}_\alpha(t)+\epsilon_\beta(t) \p_\beta v_\alpha(\bx)\,, &\quad 
 \delta p(\bx) =\epsilon_\beta(t) \p_\beta p(\bx)\,, \nonumber\\
 \delta \bar v_\alpha(\bx) =\epsilon_\beta(t) \p_\beta \bar
 v_\alpha(\bx)\,, &\quad \delta \bar p(\bx) = \epsilon_\beta(t) \p_\beta \bar p(\bx) \, ,
\label{eq:defGal}
\end{align}
where $\vec \epsilon(t)$ is an infinitesimal function of time, and $\dot{\epsilon}_\alpha = \p_t \epsilon_\alpha$. 
  The special case $\vec\epsilon(t)\equiv \vec \epsilon$ corresponds to a translation in space  and $\vec\epsilon(t)\equiv \vec \epsilon\;t$ to the usual (non-gauged) Galilean symmetry.
 This transformation is an extended symmetry in the sense that the \NS action is not strictly invariant under the transformation \eq{eq:defGal},
 but  has a variation linear  in the fields, and this can be exploited to derive useful Ward identities, summarized in \citep{Canet15}. One proceeds as follows.
 A functional Ward identity is obtained by explicitly performing the transformation \eq{eq:defGal} as a change of variables in the 
 functional integral (\ref{eq:Z}), and noting that the measure is left unchanged.  
 This functional identity reads for $\Gamma_\kappa$, (omitting pressure terms which play no role  in what follows, see \citep{Canet16} for the explicit treatment of the pressure terms) 
\begin{equation}
\label{eq:wardGal}
 \int_{\vx} \Big\{\big(\delta_{\alpha\beta}\p_t +  \p_\beta 
u_\alpha\big)\frac{\delta \Gamma_\kappa}{\delta u_\alpha}   \nonumber
+ \p_\beta \bar u_\alpha \frac{\delta \Gamma_\kappa}{\delta \bar 
u_\alpha} \Big\}= -\int_{\vx} \p_t^2 \bar u_\beta \, .
\end{equation}
By taking functional derivatives of this identity with respect to velocity and response velocity fields, and setting the fields to zero, one can derive  exact identities for  (\PI) generalized correlation functions which reads (see \aref{AP:Ward}):
\begin{align}
  &\bar\Gamma^{(m+1,n)}_{\alpha\alpha_1\dots\alpha_{n+m}}(\omega,\vp=0,\bp_1, \dots , \bp_{n+m-1})\nonumber\\
  &= -\sum_{k=1}^{n+m-1}\frac{p_k^{\alpha}}{\omega} \Big[\bar\Gamma^{(m,n)}_{\alpha_1\dots\alpha_{n+m}}(\overbrace{\bp_i}^{k-1},\,\nu_k+\omega,\vp_k,\overbrace{\bp_j}^{m+n-k-1}) \nonumber\\
  &\quad\quad\quad\quad - \bar \Gamma^{(m,n)}_{\alpha_1\dots\alpha_{n+m}}(\underbrace{\bp_i}_{m+n-1})  \Big]\nonumber\\
 &\equiv {\cal D}_\alpha(\omega) \bar \Gamma^{(m,n)}_{\alpha_1\dots\alpha_{n+m}}(\bp_1, \dots, \bp_{n+m-1}) \, . \label{eq:wardGalN}
\end{align}
The operator ${\cal D}_\alpha(\omega)$ is hence a finite difference operator, which successively shifts by $\omega$ all the frequencies of the function on which it acts.
The identities  \eq{eq:wardGalN} exactly relate an arbitrary  $(m+1,n)$-point vertex  function with one vanishing wave-vector carried by a velocity field $u_\alpha$  to a lower-order $(m,n)$-point vertex function. 

Another extended symmetry  of the \NS action, related to a time-gauged shift in the response field sector, was identified in \citep{Canet15}. It corresponds to the field transformation 
\begin{equation}
 \delta \bar v_\alpha(\bx) =\bar \epsilon_\alpha(t)\,,\quad
 \delta \bar p(\bx)= v_\beta(\bx) \bar \epsilon_\beta(t) \, .\label{eq:defShift}
\end{equation}
Writing that this change of variables leaves the functional integral \eq{eq:Z} unchanged leads to the following functional
Ward identity \citep{Canet15}:
\begin{equation}
\label{eq:wardShift}
\int_{\vx} \Big\{\frac{\delta \Gamma_\kappa}{\delta \bar u_\beta}
 + u_\beta \p_\gamma u_\gamma \Big\}= \int_{\vx}  \p_t u_\beta \, .
\end{equation}
Taking functional derivatives with respect to  velocity and response velocity fields and evaluating at zero fields, one can deduce again exact identities for  vertex functions \citep{Canet16}.  They give the expression of any $\bar\Gamma^{(m,n)}$ with one vanishing wave-vector carried by a response velocity, which reads
\begin{align}
 &\bar\Gamma_{\alpha_1\dots\alpha_{n+m}}^{(m,n)}(\omega_1,\vec p_1, 
\dots,\omega_m,\vec p_m, \omega_{m+1},\vec p_{m+1}=\vec 0,\dots)= 0 \, ,
  \label{eq:wardShiftN}
\end{align}
 for all $(m,n)$ except for the two lower-order ones which keep their bare form:
\begin{align}
\bar \Gamma_{\alpha\beta}^{(1,1)}(\omega,\vec p=\vec 
0)&=i\omega\delta_{\alpha\beta},\nonumber\\
\bar\Gamma_{\alpha\beta\gamma}^{(2,1)}(\omega_1,\vec p_1,\omega_2,-\vec p_1) &= i 
p_1^\alpha \delta_{\beta\gamma} -ip_1^\beta \delta_{\alpha\gamma} \label{eq:wardShift21}\, .
\end{align}
Note that some identities that may be related to these symmetries were derived
within the quasi-Lagrangian framework \cite{Lvov93}. However, the underlying symmetry was not identified. All
these Ward identities are exploited in the next section to
achieve a  closure of the flow equation for a generic
correlation function which corresponds to the exact leading contribution in the limit of large wave-numbers.

\section{Flow equation for generic correlation functions}
\label{secFlow}

The aim of this section is to show that the  flow equation for a generic generalized connected correlation function $\bar G^{(n)}$
 can be closed, \ie expressed in terms of  $\bar G^{(k)}$ with $k\leq n$ only, through a controlled  expansion in the limit of 
  large wave-numbers  with respect to the \RG scale $\kappa$, and for non-exceptional wave-vector configurations.
 This derivation relies on two key ingredients.  The first ingredient is the Blaizot-M\'endez-Wschebor (BMW) approximation 
 \citep{Blaizot06},  which consists, when external momenta are large, in expanding the $\Gamma^{(n)}$ vertices in
 the flow equation in series of $|\vq|/|\vp_i|$ where $\vq$ is the internal wave-vector and $\vp_i$
  is some external one. Accordingly, this approximation
  becomes exact when  all wave-vectors (and all their partial sums) are large with respect to $\kappa$.
 This approximation relies on the presence of the regulator term $\p_\kappa {\cal R}_\kappa(\vq)$
 in the flow equation of $\bar G^{(n)}$. As emphasized with \Eq{eq:limitreg}, this function rapidly vanishes for wave-numbers 
 larger than the \RG scale $\kappa$.
 On the other hand, the regulator ensures the analyticity of vertex functions at any finite $\kappa$. In particular, 
 if a vertex function in the flow equation depends on external wave-vectors $\vp_i$ and on the internal wave-vector 
 $\vq$, then the latter is negligible in the limit of large wave-numbers $|\vp_i| \gg \kappa$ since it is cut off by 
 $\p_\kappa {\cal R}_\kappa$ to values $|\vq| \lesssim \kappa$. 
 As a consequence, this wave-vector can be safely set to zero within the vertex function
 because of the regularization
 \footnote{Let us stress that such an approximation can only be applied to \PI vertices. Indeed, for connected correlation functions, the wave-number $|\vq|$ can not always  be compared to the external wave-numbers $|\vp_i|$ because of the presence of propagators of the type $G^{(2)}(\vq)$. This does not happen for \PI vertices.}.

 The second key ingredient is the set of Ward identities \eq{eq:wardGalN}, \eq{eq:wardShift21} and \eq{eq:wardShiftN}, which exactly fixes the expression of vertex functions with one vanishing wave-vector  in terms of lower-order ones.
 Hence, the principles of the derivation is to combine both ingredients to close the flow equation for any  $\bar G^{(n)}$. This can be successfully achieved because both the velocity sector and response velocity sector are constrained by respectively the extended Galilean and extended shift symmetries.

 Let us already emphasize a very unusual feature of the flow equations obtained in this paper. The large wave-number part of the flow equation (determined  exactly using the BMW framework) is not negligible compared to the rest of the flow. This behavior is referred to as non-decoupling \citep{Canet16,Canet17}.
 It implies that the \RG flow of the UV modes depends on the IR ones. This is in sharp contrast with what occurs in standard critical phenomena, where the large wave-number part of the flow equation decouples from the small wave-number one, and  this decoupling precisely entails standard scale invariance. On the contrary, the non-decoupling induces a violation of standard scale invariance, as manifest in the following.

 The explicit derivation of  the flow equation for a generic function $\bar G^{(n)}$ is detailed in \aref{AP:flowN}.
 The starting point is the exact equation   obtained by taking $n$ functional derivatives of \Eq{eq:dkw} with respects to 
 the appropriate sources $j_i$.  One can first establish that only the following contribution survives in its flow equation
 in the limit of large wave-numbers and at zero fields:
\begin{align}
\p_\kappa & \bar G^{(n)}_{\alpha_1\dots\alpha_{n}}({\bp_1}, \dots, {\bp_{n-1}}) = 
 \frac{1}{2}\,  \int_{\bq}\! \tilde \p_\kappa \bar G^{(2,0)}_{\mu \nu}(\bq) \nonumber\\
 &\times \left[ \frac{\delta^2}{\delta u_\mu(\bq)\delta u_\nu(-\bq)} \bar G^{(n)}_{\alpha_1\dots\alpha_{n}}[{\bp_1}, \dots, {\bp_{n-1}};j] \right]_{\varphi=0} \, ,
\end{align}
where the dependency of $\bar G^{(n)}$ in $u$ appears through the implicit dependency $j=j[\varphi]$ and with the differential operator $\tilde \p_\kappa$ defined as:
\begin{equation}
 \tilde \p_\kappa \equiv \p_\kappa R_\kappa \frac{\delta}{\delta R_\kappa} + \p_\kappa N_\kappa \frac{\delta}{\delta N_\kappa} \, .
\end{equation}

 The second step is to show the following property: taking  two functional derivatives with respect to a velocity field $u_\alpha(\omega,\vq)$  carrying the internal wave-vector and evaluating at vanishing fields is equivalent in the limit of large wave-numbers to applying twice the discrete operator ${\cal D}_\alpha(\omega)$ defined in \eq{eq:wardGalN}. 
 One then deduces  that the leading contribution of the flow equation for large wave-numbers reads
\begin{align}
\p_\kappa &\bar G^{(n)}_{\alpha_1\dots\alpha_{n}}({\bp_1}, \dots, {\bp_{n-1}}) = 
 \frac{d-1}{2d}\,  \int_{\bq}\! \tilde \p_\kappa \bar C(\varpi,\vq) \nonumber\\
  & \times {\cal D}_\mu(\varpi){\cal D}_\mu(-\varpi) \bar G^{(n)}_{\alpha_1\dots\alpha_{n}}({\bp_1}, \dots, {\bp_{n-1}})\, ,
\label{eq:flowGN}
\end{align}
where $\bar C$ is the transverse part of the velocity-velocity correlation function $\bar G_{\alpha_1\alpha_2}^{(2,0)}$.
The flow equation for any generalized correlation function $\bar G^{(n)}$ is hence closed, in the sense 
that it does not depend any longer on higher-order correlation functions. This closure is exact in the limit
 of large wave-numbers, when the modulus of all wave-vectors and  of all their partial sums are large compared
  to $\kappa$, which excludes exceptional configurations where a partial sum vanishes. We emphasize that this closure
   involves no arbitrary truncation or selection of certain diagrams rather than others. Its rationale is
    to retain only
    the leading order contribution in wave-number in the flow equation, and this contribution is calculated exactly.
  This constitutes the first central result of the paper. Note that the second line in \Eq{eq:flowGN} does not 
  depend on the internal wave-vector $\vq$, and is only integrated over the internal frequency $\varpi$. 
 This result generalizes a previous result obtained in \citep{Canet16} for $2-$point functions to generic $n-$point
  connected functions. In the next sections, we study some aspects of the solution of this flow equation at the fixed point, 
  and in particular the general form of the space and time dependence of the correlation functions. 

\section{Solution for the 2-point functions}
\label{sec2point}

The leading contribution of the flow equation for the 2-point functions in the limit of large wave-numbers was already derived in \citep{Canet16}.
In the notation of the present paper, \Eq{eq:flowGN}, it reads
\begin{align}
 \p_\kappa \bar G^{(2)}_{\alpha_1\alpha_{2}}({\bp}) =  \frac{d-1}{2d}\,  &\int_{\varpi}\! {\cal D}_\mu(\varpi){\cal D}_\mu(-\varpi)  \bar G^{(2)}_{\alpha_1\alpha_{2}}({\bp}) \nonumber\\
 &\times \int_{\vq}\tilde \p_\kappa \bar C(\varpi,\vq) \, .
\end{align}
This flow equation encompasses both the flow of the correlation function and of the response function. Let us denote $\bar C$ and $\bar G$ their respective transverse part,
that is  $\bar G^{(2,0)}_{\mu\nu}({\bp}) =P_{\mu\nu}^\perp(\vp) \bar C({\bp})$ and $\bar G^{(1,1)}_{\mu\nu}({\bp}) =P_{\mu\nu}^\perp(\vp) \bar G({\bp})$,
 where the transverse projector is defined by
\begin{equation}
 P_{\mu\nu}^\perp(\vp)=\delta_{\mu\nu}-\frac{p_\mu p_\nu}{p^2}\, .
\end{equation}
Let us focus on the flow equation for the transverse velocity-velocity correlation function. Using the explicit expression \eq{eq:wardGalN} for ${\cal D}_\mu$, 
 one obtains:
\begin{equation}
\kappa \p_\kappa \bar C(\omega, \vp)=-  \frac{2}{3} p^2 \int_{\varpi}
\frac{\bar C(\omega+\varpi, \vp)- \bar C(\omega, \vp)}{\varpi^2} \, J_\kappa(\varpi) \, ,
\label{eq:flowC}
\end{equation}
  with $J_\kappa(\varpi) \equiv -\int_\vq \tilde \p_\kappa \bar C(\varpi,\vq)$  given by
 \begin{align}
J_\kappa(\varpi)
  =&- 2  \,\int_{\vq} \Bigg\{ \kappa \p_\kappa N_\kappa(\vq) \,|\bar G(\varpi,\vq)|^2 \nonumber\\
  &-  \kappa \p_\kappa R_\kappa(\vq)\,\bar C(\varpi,\vq) \Re \big[\bar G(\varpi,\vq)\big]\Bigg\}\, ,
 \label{eq:Ikomega}
\end{align}
(which coincides with the equations given in \cite{Canet17}).
An important feature of \Eq{eq:flowC},  already emphasized,  is the non-decoupling, which means that 
$\kappa \p_\kappa \bar C/\bar C$ does not vanish in the limit of large wave-numbers $|\vp| \gg \kappa$.
 It was shown in \citep{Canet16} that this property implies the violation of standard scale invariance.
 As a consequence, the behavior of the correlation functions  at $t \neq 0$  shows non-standard scale invariance. 
  On the other hand, at equal times, that is once integrated over the external frequency $\omega$, the leading 
  non-decoupling behavior cancels out (the r.h.s. of \Eq{eq:flowC} vanishes when integrated over $\omega$). For equal-times quantities, any possible non-decoupling must come from 
  sub-leading terms at large wave-numbers. This implies in particular that intermittency corrections to exponents 
  of equal-time quantities, such as the second order structure function, are absent in the leading order behavior
   at large $|\vp|$, \ie not included in these flow equations.
 This is made explicit in the solution below.

 In the next sections, we derive the solutions of the flow equation \eq{eq:flowC} at the fixed point.
  It is convenient to first  perform the inverse Fourier transform on $\omega$, which yields
\begin{equation}
\kappa \partial_\kappa  C(t, \vp)=-\frac{2}{3}  p^2 C(t,\vp) \int_{\varpi} 
 \frac{\cos(\varpi t) -1}{\varpi^2}\, J_\kappa(\varpi)\, .
\label{eq:flowCt}
\end{equation}
The regimes of small $t$ and large $t$ are studied separatly in the following.

 \subsection{Small time delays}

The  solution of \Eq{eq:flowCt} at the fixed point and in the limit of small time delays (or equivalently large frequencies) has been obtained in \citep{Canet17}. For small $t$, $(\cos(\varpi t) -1)/\varpi^2\sim -t^2/2$ in the integrand.
 The general solution of the resulting fixed-point equation,     
 which we index with the subscript $'$S$'$ for 'short time',  follows:
\begin{align}
 \log \Big[\frac{C_S(t,\vp)}{\varepsilon^{2/3}L^{11/3}} \Big] &= - \alpha_{S}(\varepsilon L)^{2/3} t^2 p^2 -\frac{11}{3} \log (pL) \nonumber\\
 &+ F_S(\varepsilon^{1/3}p^{2/3}t) + {\cal O}(pL) \,, 
 \label{eq:solC}
\end{align}
where $\varepsilon$ is the mean energy injection rate,  $L$ the integral scale,  $\alpha_S$ 
 a non-universal  constant depending on the forcing profile (see \aref{AP:solution}), and  $F_S$ is a regular function, 
  universal up to a pre-factor. 
   Note that we have included explicitly in \eq{eq:solC} sub-leading terms stemming from the resolution of the fixed-point
 equation at leading order, although they are of the same order
   as the error term ${\cal O}(pL)$. The reason is that they correspond to Kolmogorov scaling solution, and facilitate
  the discussion of the result. Indeed,
   at equal-time, one recovers from  \eq{eq:solC} the Kolmogorov prediction
   $C_S(t=0,\vp) = H_S(0) \varepsilon^{2/3} p^{-11/3}$. However, 
    these terms are only approximate  in this calculation. If the sub-leading terms 
  in the flow equation \eq{eq:flowCt}, neglected here, are non-decoupling, 
   then they will induce corrections (of order $pL$) to this scaling, that is,  intermittency corrections 
 to the exponent of the structure function. 
    The calculation of the sub-leading terms of   \eq{eq:flowCt} is left for future investigation.

 On the other hand, at finite time delays $t \neq 0$, the leading term in \eq{eq:solC}
  explicitly breaks scale invariance. Indeed, it does not depend on the scaling variable
   $tp^z$, where $z$ is the
  dynamical exponent   $z=2/3$ for \NS in $d=3$, and thus involves a scale $L$.
    This leading term
 conveys  an effective exponent $z=1$, which indicates significant corrections to standard scale invariance.
 Its physical origin
   is the sweeping effect, which is the random advection of small-scale velocities by large-scale eddies 
   \citep{Tennekes75,Yakhot89,Chen89,Nelkin89,Gotoh93,Chevillard05}. The typical time-scale appearing in the exponential \eq{eq:solC} is the sweeping time $\tau_S\sim (\varepsilon L)^{-1/3}p^{-1} = 1/(u_{\rm rms}p)$.
 One of its manifestation is that the spectrum measured in frequency has the same exponent -5/3 as when measured in
  wave-numbers. 
 (We consider flows with zero mean velocity, this is not related to  Taylor's hypothesis of frozen turbulence).

The behavior \eq{eq:solC}  has been observed in many numerical simulations of the \NS equation \citep{Orszag72,Sanada92,He04,Favier10,Canet17} as well as in experiments \citep{Poulain06}.
 Indeed, in Fig. 1 of ~\cite{Orszag72}, and Fig. 5 of ~\cite{He04}, a reasonable collapse is obtained for the quantity $R(t,p) = C(t,p)/C(0,p)$ as a function of $p\int_t u_{rms}$ for
 different times and different values of wave-numbers respectively, from simulations of decaying turbulence. In Fig. 6 and 7 of~\cite{Sanada92}, the collapse for the same quantity as a function
 of time is qualitatively better when normalizing by the sweeping time $\tau_S$ rather than the eddy turn-over time $\tau_e \sim 1/(\varepsilon ^{1/3}p^{2/3})$.
In Fig. 7, the typical time scale of $R(t,p)$ is shown to scale linearly in $p$ for large wave-numbers. The same analysis is carried out  in Fig. 4 of~\cite{Favier10}. The  time dependence of the two-point function is explicitly tested in~\cite{Canet17}, where a Gaussian form of $R(t,p)$  in the variable $u_{rms}\,p\,t$  is very accurately found in numerical simulations. The Gaussian behavior and the linear dependency in $p$ of the decorrelation time of $R(t,p)$ is also found in acoustic
scattering measurements, see Fig. 5 and 6 in~\cite{Poulain06}.
 As mentioned in the introduction, such a Gaussian dependence in $tp$ for large $p$ and small $t$ was
  predicted early on by Kraichnan within the DIA approximation \citep{Kraichnan59}, and later confirmed by \RG approaches under some assumptions on  the effective viscosity \cite{Antonov94}.

Let us briefly mention  another feature of the fixed point solution of \Eq{eq:flowC}.
   It was shown   in \cite{Canet17} that, 
   under some additional assumptions,  taking the appropriate $t\to 0$ limit,
    this solution predicts for the kinetic energy spectrum,
   a crossover from the $p^{-5/3}$ decay in the inertial range, to a stretched exponential decay 
    in the dissipative range, on the scale $p^{2/3}$
 \begin{equation}
E(\vp) \propto  p^{-5/3}   \exp\left[- \mu  p^{2/3}  \right]  \, ,
\label{eq:exp}
\end{equation}
with $\mu$ a non-universal constant.
This prediction was  precisely confirmed in direct numerical simulation of \NS equation \cite{Canet17}, 
and also observed  in experiments on turbulent swirling flows \cite{Debue17}.

 \subsection{Large time delays}
\label{sec:C2largeT}

 The flow equation \eq{eq:flowCt} is valid for a large wave-number $p$, but for an arbitrary time
  delay $t$. In this section, we study the opposite limit of asymptotically large $t$, which was not considered previously.
 As shown in \aref{AP:solution}, the flow equation \eq{eq:flowCt}  simplifies in the limit $t\gg \kappa^{2/3}$ to
\begin{equation}
\kappa \partial_\kappa  C(t, \vp)= \frac{J_\kappa(0)}{3} \,|t|\, p^2 \,C(t,\vp). \label{eq:flowClargeT}
\end{equation}
As for the flow equation at small $t$, this equation can be  solved at the fixed point (see \aref{AP:solution}).
  The solution, indexed by the subscript '$L$' for 'long' time, reads 
\begin{align}
\log \Big[\frac{ C_L(t,\vp)}{\varepsilon^{2/3} L^{11/3}}\Big] &= - \alpha_{L} \varepsilon^{1/3} L^{4/3} |t|\, p^2
 -\frac{11}{3}\log (pL) \nonumber\\
  &+ F_L(\varepsilon^{1/3}k^{2/3}t) + {\cal O}(pL)\, , \label{eq:solCL}
\end{align}
with $\alpha_{L}$  a non-universal constant, and sub-leading terms corresponding to Kolmogorov
 solution again included explicitly.
To the best of our knowledge, this regime was not predicted before. 
The corresponding time-scale in the exponential is
 $\tau_L = (u_{\rm rms}L p^2)^{-1}$.

Interestingly, a similar crossover from the Gaussian in $tp$ at short time to a behavior $\exp(-|t|/\tau_{\rm exp})$ at long times was observed  in \cite{Poulain06}. However, in this paper,
   $\tau_{\rm exp}  \propto (u_{\rm rms})^{-1}L$, that is the $p^2$ in $\tau_L$ is replaced by  $L^{-2}$.
This indicates that the crossover seen in the experiments is dominated by the small wave-numbers, so it is likely to differ from ours.
 One can compare the related time-scales of these two crossovers. 
 The crossover  between the two (short and long time) regimes occurs typically when the exponents in the two exponentials are equal.
For our work, matching the exponents in  \eq{eq:solC} and  \eq{eq:solCL} 
   yields $\tau_{\rm cross} \propto L$.
In the experimental paper, the crossover time is given by $\tau_{\rm cross} \propto L^2/p=L/(p L^{-1})$. Hence, at large $p \gg L^{-1}$, this crossover time is shorter than the second crossover, and may dominate over it.

\section{Form of the solution for generic correlation functions}
\label{secSol}

In this section, we work out the general form of the fixed point solution of the flow equation \eq{eq:flowGN} for any  generalized  $n-$point correlation functions.
From equation \eq{eq:flowGN}, the first step is to perform the inverse Fourier transform in frequency in order to get the flow equation
for the hybrid wave-vector and time correlation functions.
One obtains  for the leading contribution of the flow equation at large wave-numbers

\begin{align}
 &\p_\kappa G^{(n)}_{\alpha_1\dots\alpha_{n}} (\{{t_i,\vp_i}\}) = \frac{1}{3}\,  G^{(n)}_{\alpha_1\dots\alpha_{n}}  (\{{t_i,\vp_i}\})\nonumber\\
&\times\sum_{k,\ell} \vp_k \cdot \vp_\ell \int_{\varpi}\! J_\kappa(\varpi)\,
\frac{e^{i\varpi(t_k - t_\ell )} - e^{i\varpi t_k} - e^{-i\varpi t_\ell} +1}{\varpi^2} \, .
\label{eq:lowGt}
\end{align}

The solution of the corresponding fixed-point equation is derived in the following in both limits
 $t\to 0$ and $t\to\infty$.

\subsection{Small time delays}

For $t_i \ll \kappa^{-2/3}$, the flow equation \eq{eq:lowGt} simplifies to (see \aref{AP:solution})
\begin{equation}
 \label{eq:flowSmallTimes}
 \Big(\p_\kappa -\frac{I_{\kappa}}{3}\, |\vp_k t_k|^2 \Big) G^{(n)}_{\alpha_1\dots\alpha_{n}}({t_1,\vp_1} ,\cdots, {t_{n-1},\vp_{n-1}}) =0\, ,
\end{equation}
with $I_{\kappa} =  \int_{\varpi} J_{\kappa}(\varpi)$, and $J_\kappa$ defined in  \Eq{eq:Ikomega}, and with Einstein summation convention. 
The corresponding fixed point equation can be solved exactly, leaving as unknown a scaling function of particular dimensionless variables. 
Let us present this solution, which is derived in details in \aref{AP:solution}, and which reads
\begin{align}
\label{eq:exprGn}
 &\log\Big[ \varepsilon^{\frac{\bar{m}-m}{3}} L^{-d_G}{{G}_S^{(n)}}_{\alpha_1\dots\alpha_{n}}({t_1,\vp_1}, 
 \cdots,t_{n-1}, {\vp_{n-1}})\Big] \nonumber\\
 &= -\alpha_S\varepsilon^{2/3} L^{2/3} \,  \abs{t_k \vp_k}^2 -d_G \log (\rho_1 L) \nonumber\\
  & + {F_S^{(n)}}_{\alpha_1\dots\alpha_{n}}\left({\varepsilon^{1/3}\rho_1^{2/3} t_1},\frac{\vec{\rho}_{1}}{\rho_1}, \cdots,
  \varepsilon^{1/3}\rho_1^{2/3}t_{n-1},{\frac{\vec{\rho}_{n-1}}{\rho_1}}\right)\nonumber\\
  &+ {\cal O}(p_{\rm max}L)\, .
\end{align}
 In this expression, $d_G$ is the scaling dimension of ${G}^{(n)}$ (given in \aref{AP:solution}),
  $m$ (resp. $\bar{m}$) is the number of velocity (resp. response velocity) fields in this generalized 
  correlation function, with $m+\bar m=n$,  and $\alpha_S$ is the same non-universal constant as in \Eq{eq:solC}. 
  $F_S^{(n)}$ is a regular  function of its arguments, which cannot be determined by the fixed point equation alone, 
  but requires the integration of the full flow equation. 
  As for the two-point functions, sub-leading terms which correspond to Kolmogorov scaling solution 
  are included in this expression, although they are only approximate and could receive corrections from the 
   neglected ${\cal O}(p_{\rm max}L)$ terms.
  The variables $\vec{\rho}_k$ are defined by $\vp_i = {\cal R}_{ij}\vec{\rho}_j$ where ${\cal R}_{ij}$ is a rotation matrix which has to be explicitly constructed  for each correlation function such that
\begin{equation}
 \vec{\rho}_1 = \frac{ t_k \vp_k}{\sqrt{t_\ell t_\ell}}\, .
\label{eq:variables}
\end{equation}
 Finally, $p_{\rm max}$ is the maximum amplitude of the $\vp_i$ and their partial sums. This is the second main 
 result of this paper. It provides  the leading time and wave-vector dependence of any correlation functions, 
  which is exact in the regime of small time differences and large wave-numbers.
The combination of time and space appearing in the exponential part of the expression \eq{eq:exprGn} is $|\vp_k t_k|^2$. This combination breaks scale invariance and is the generalization to generic $n$-point correlation functions of the Gaussian dependence in  the variable $(p\, t)$ for the $2-$point correlation function, which is related to the sweeping effect.
 This breaking yields the dependence on the integral scale $L$ of this leading term. 
 Furthermore, in the range of validity of this solution, one has $p^{2/3} t \ll p\, t\, L^{1/3}$. Hence, because 
 of the regularity of $F_S^{(n)}$, the leading time contribution is due in this regime to the variable $\vp_k t_k$,
  except for exceptional configuration where $\vp_k t_k \simeq 0$.
 The leading time-dependence of the correlation function hence takes  the form of a Gaussian
\begin{equation}
 {G}_S^{(n)} \sim e^{-\alpha_S(\varepsilon L)^{2/3}|\vp_k t_k|^2}\, .
\end{equation}
If wave-vectors are measured in units of $\eta^{-1}$ as is usually the case,  the resulting typical time scale  is the sweeping time  $\tau_{\rm s} = \eta/u_{\rm rms} = \eta/(\varepsilon L)^{1/3}$, which differs from the Kolmogorov time $\tau_{K}=(\nu/\varepsilon)^{1/2}$.

 On the other hand, at equal times, one is left with
  \begin{align}
  {{G}_S^{(n)}}_{\alpha_1\dots\alpha_{n}}(\{0,\vp_i\}) &= \varepsilon^{\frac{m-\bar{m}}{3}}{\rho_1^{-d_G}} \,\nonumber\\
   &\times \exp {F_S^{(n)}}_{\alpha_1\dots\alpha_{n}}\left(\left\{0,\frac{\vec{\rho}_{i}}{\rho_1}\right\}\right)
  \end{align}
  which corresponds to  Kolmogorov-like  solution : a power-law behavior with a dimensional
   exponent times a scaling function. However, our calculation shows that this is not exact {\it a priori}, 
   and is thus compatible with the existence of intermittency corrections. 
   Indeed,  these terms can receive corrections from the  neglected ${\cal O}(p_{\rm max}L)$ terms in the flow equation.
    This terms  could in particular modify the exponent of the power-law, that is yield intermittency correction 
    to the structure functions. These corrections should be given by the next order term 
    in the flow equation, provided it does not to vanish at equal time. This is work in progress.

We now specialize  to the three-velocity correlation. In the regime of small time differences and large wave-numbers, we obtain
\begin{align}
 {{G}_S^{(3)}}_{\alpha\beta\gamma}({t_1,\vp_1}, {t_{2},\vp_{2}}) &\sim {{G}_S^{(3)}}_{\alpha\beta\gamma}({0,\vp_1}, {0,\vp_{2}}) \nonumber\\
 & \times \exp\Big(-\alpha_S(\varepsilon L)^{2/3}  \abs{\vp_1t_1 + \vp_2t_2}^2\Big)\,.
\end{align}
This prediction can be tested in direct numerical simulations of the \NS equation or in experiments.
For example, one can  construct a scalar function from the 3-velocity correlation, such as $p_1^\alpha G_{\beta\beta \alpha}^{(3,0)}(\vp_1,t,\vp_2,t)$, and measure its dependence in the time difference $t$
 in the stationary state. Normalizing the constructed  function by its value at $t=0$, one should obtain a Gaussian dependence in the variable $\abs{\vp_1 + \vp_2}t$.

\subsection{Large time delays}

We consider again the flow equation \Eq{eq:lowGt}.
In the limit of large times, \ie all times $t_k \gg \kappa^{2/3}$ as well as all differences $(t_k-t_\ell) \gg \kappa^{2/3}$, 
 this equation   simplifies to (see \aref{AP:solution})
\begin{align}
 &\p_\kappa G^{(n)}_{\alpha_1\dots\alpha_{n}}(\{t_i,\vp_i\})=\frac{J_{\kappa}(0)}{6}\,\nonumber\\
&  \times \sum_{k,\ell} \vp_k\cdot\vp_\ell \Big( |t_k| +|t_\ell| -|t_k-t_\ell| \Big)\bar G^{(n)}_{\alpha_1\dots\alpha_{n}}(\{t_i,\vp_i\}) \, .
\end{align}
We  focus on the special case where all the time delays are equal
$t_k\equiv t$ for $k=1,\dots,n-1$. In this case,  the analytical solution of the corresponding fixed-point equation can be straightforwardly  derived (see \aref{AP:solution}). 
One obtains, keeping again the sub-leading Kolmogorov scaling terms, 
\begin{align}
 &\log \Big[\varepsilon^{\frac{\bar{m}-m}{3}}L^{-d_G} {G}^{(n)}_{\alpha_1\dots\alpha_{n}}(t,{\vp_1}, \cdots, 
 {\vp_{n-1}})\Big]\nonumber\\
 &= -\alpha_L\varepsilon^{1/3}L^{4/3} \,  |t|\, \left|\sum \vp_k\right|^2 -d_G \log (\varrho_1 L)\nonumber \\
 & +{{F_L}^{(n)}}_{\alpha_1\dots\alpha_{n}}\left(\varrho_1^{2/3}\varepsilon^{1/3} t,\frac{\vec{\varrho}_{1}}{\varrho_1}, \cdots, \frac{\vec{\varrho}_{n-1}}{\varrho_1}\right)\,\nonumber\\
 &+{\cal O}(p_{\rm max}L)\, ,
\end{align}
where the variables $\vec \varrho_k$ are obtained by a transformation of the wave-vectors satisfying  
 $\vec \varrho_1 =\sum \vp_k$, which can be explicitly constructed for each $n$. The example of $G^{(3)}$ is explicitly given in \aref{AP:solution}.
 The crossover, evidenced for the two-point function, also emerges for generic $n$-point functions. The quadratic dependence in $t$ in the exponential at small time delays is changed at large time delays to a linear one.
 This regime, which we believe was not predicted before, is the last  result of the paper.\\

\section{Conclusion and perspectives}

In this paper,  we have derived theoretical results related to the statistical properties of 
 fully developed isotropic and homogeneous turbulence in three dimensions. 
 The first result is a closed  flow equation for any generalized $n$-point correlation  functions, encompassing their 
  space and time dependence. The closure we achieve, which relies on the existence of extended symmetries of the \NS 
  field theory, 
 is based on an expansion at large wave-numbers. In this work, we calculate exactly the leading term in this expansion,
 for non-exceptional wave-vector configurations (none of the partial sums of wave-vectors vanish).

The second result is the analytical form of the fixed-point solution of this equation both in the limit of small and large 
 time delays.
 The fixed point equation is appropriate to describe the universal properties of stationary turbulence.
 At small time delays, we find that all the correlation functions exhibit the same specific leading behavior
  $\propto \exp(-\alpha |\sum t_i \vp_i|^2)$,  as a generalization of the sweeping effect. We emphasize
   that the correlation functions do not endow the usual dynamical
 scaling form in terms of the scaling variables $t p^z$, which indicates a
  violation of standard scale invariance, and thus of K41 theory.
 At large time delays, we find a crossover to a $\abs{ \sum \vp_i}^2 \abs{t}$ dependence
  in the exponential (for all time delays equal). This also violates standard scale invariance, 
  but seems to originate from a different mechanism.
  All these predictions can be tested in direct numerical simulations or in experiments.

The outcome of our calculation is an analytical expression of the correlation functions,
 which is asymptotically exact at large wave-numbers. However, this expression  does not include
  intermittency corrections to the exponents of the structure functions, since we have shown that intermittency effects are not present for equal-time quantities  at leading order in wave-numbers. 
  The calculation of the sub-leading terms in the present scheme, or using alternative approaches,
   combining a suitable approximation for small wave-numbers (similar to the ones developed e.g. in
    \citep{Canet11a,Kloss12,Canet16}) with the present one, are required to compute structure functions 
    and determine the associated intermittency corrections to their scaling exponents. This is left for future work.\\

\begin{acknowledgments}
The authors thank B. Delamotte and C. Pagani for fruitful discussions, and the latter
 for the suggestion to study the large time behavior. N. W. thanks Pedeciba (Programa de desarrollo de las Ciencias B\'asicas, Uruguay)
 and acknowledges funding through  grant from la Comisi\'on Sectorial de Investigaci\'on Cient\'ifica de la Universidad de la Rep\'ublica, Project I+D 2016 (cod 412).
\end{acknowledgments}

\appendix

\begin{widetext}

\section{MSRJD response functional formalism}
\label{AP:MSRJD}

In order to use field theoretical techniques to study Navier-Stokes equation, one needs to formulate the associated 
 field theory. The MSRJD response functional formalism is a standard procedure to turn a classical Langevin equation into a field theory.
 Thus, one usually considers, as done in \sref{secNS}, a stochastic force (concentrated at the integral scale) in the  Navier-Stokes equation
  and then apply the MSRJD formalism. The stochastic nature of the forcing is not expected to affect the properties of the turbulent flow at scales sufficiently smaller than the integral scale because of universality.  

 Let us describe the MSRJD formalism considering a set of $n$ generic stochastic fields $\Phi_i(\bx)$ defined through a set of $m$ stochastic partial differential  equations with $n$ constraints
\begin{align}
 \p_t \Phi_i &= F_i(\Phi) + G_{ij}(\Phi)\xi_j\,,\quad1\leq i \leq m\nonumber\\
 \,H_i(\Phi) &= 0\,,\quad 1\leq i \leq n
 \label{eq:SPDE}
\end{align}
where the $F_i$, $G_{ij}$ and $H_i$ are functions of the $\Phi_i$ and their spatial derivatives and the $\xi_i$ are centered stationary Gaussian fields of correlator $D_{ij}$, 
that is, they satisfy for given fields $j_i$
\begin{equation}
 \mean{e^{\int_{\bx} j_i \xi_i}} = e^{\frac{1}{2} \int_{\bx,\bx'} j_i(\bx) D_{ij}(\bx-\bx') j_j(\bx')}\,.
 \label{eq:xi}
\end{equation}
For the Navier-Stokes equation, one has respectively
\begin{align}
 F_\alpha &= - v_\beta \p_\beta v_\alpha + \nu \nabla^2 v_\alpha - \rho^{-1} \p_\alpha p\,,\nonumber\\
 G_{\alpha\beta} &= \delta_{\alpha\beta}\,,\nonumber\\
 H &= \p_\alpha v_\alpha\,,\nonumber\\
 D_{\alpha\beta}(\bx-\bx') &= 2\, \delta_{\alpha\beta}\, \delta(t-t')N_{L^{-1}}(|\vx-\vx'|)\,.
\end{align}

In order to obtain the generating functional of the $\Phi_i$, as well as their dynamical
responses, one introduces linear source terms $\bar j_i, \bar k_i$ to the right-hand sides of equations (\ref{eq:SPDE}). The corresponding generalized generating functional reads
\begin{equation}
 {\cal Z}[j,\bar j, \bar k] = \mean{e^{\int_{\bx} j_i \Phi_i}}_{\bar j, \bar k}\,,
\end{equation}
where $\mean{\cdot}_{\bar j, \bar k}$ denotes a mean on the stochastic equation in the presence of the sources $\bar j_i, \bar k_i$. Let us now briefly explain how $\cal Z$ can be expressed as a field theory. The first step is to write $\cal Z$ as a functional integral
\begin{equation}
 {\cal Z}[j,\bar j, \bar k] = \int \mathcal D[\phi] e^{\int_{\bx} j_i \phi_i} \delta[H(\phi)-\bar k]\mean{\delta[\phi - \Phi_{\xi}]}_{\bar j}\,.
\end{equation}
The integration measure $\mathcal D[\cdot]$ and the functional Dirac delta $\delta[\cdot]$ are to be understood as the formal continuum 
limit of their discretized versions in space and time.  $\Phi_\xi$ is the weak solution of (\ref{eq:SPDE}) for a given $\xi$. The second step is to replace the constraint $\mean{\delta[\phi - \Phi_{\xi}]}_{\bar j}$ by the explicit equation of motion of $\Phi$ in the presence of the sources $\bar j_i$, which can be written as 
$\mathcal{F}(\cdot) = 0$, with
\begin{equation}
 \mathcal{F}_i(\bx) = \p_t \Phi_i - F_i(\Phi) - G_{ij}(\Phi)\xi_j - \bar j_i\,.
\end{equation}
Assume existence and unicity of weak solutions of (\ref{eq:SPDE}),  one obtains
\begin{equation}
 {\cal Z}[j,\bar j, \bar k] = \int \mathcal D[\phi] e^{\int_{\bx} j_i \phi_i} \delta[H(\phi) - \bar k]\mean{\delta[\mathcal{F}]} \times \mathcal{J}.
\end{equation}
with $\mathcal{J}$ the Jacobian of the transformation,
 $\mathcal{J} = |\det \big(\frac{\delta \mathcal{F}_i(\bx)}{\delta \phi_j(\bx')}\big)|$,
which depends on the choice of discretization of (\ref{eq:SPDE}). In this work, we  use the It\^o convention,
which amounts to have an explicit discretization scheme \footnote{Note that the  It\^o  convention is equivalent in the formal continuum limit to set the value of the Heaviside step function at
the origin to zero: $\Theta(0)=0$. In a diagramatic representation, it implies that all closed loops vanish.}.
 As a consequence, $\mathcal{J}$ can be shown not to depend on the fields~\cite{Tauber14}.
Note that  the standard derivation of the MSRJD action for \NS implicitly assumes
 existence and unicity of weak solutions of the \NS equations in three dimensions.
 This is a delicate point from a mathematical point of view, and in fact uniqueness has been shown not to hold in some cases,
  see \eg \cite{Buckmaster17}.
  However, the assumption underlying  the MSRJD derivation is a little weaker than
   strict uniqueness, since for a typical set of initial conditions, there may exist a set of velocity configurations 
   spoiling unicity, as long as they are of zero measure.

The last step is to use the Fourier representation of the functional Dirac deltas.
\begin{align}
 {\cal Z}[j,\bar j, \bar k] &= \int \mathcal D[\phi,\bar \phi,\bar h] e^{\int_{\bx} j_i \phi_i} e^{-i \int_{\bx} \bar h_i (H_i - \bar k_i)} \mean{e^{ - i \int_{\bx} \bar \phi_i \mathcal{F}_i}} \nonumber\\
 &= \int \mathcal D[\phi,\bar \phi,\bar h] e^{\int_{\bx} \{j_i \phi_i + i \bar j_i \bar \phi_i + i \bar k_i \bar h_i\}} e^{-i \int_{\bx} \bar h_i H_i} e^{ - i \int_{\bx} \bar \phi_i (\p_t \phi_i - F_i)} \mean{e^{ i \int_{\bx} \bar \phi_i G_{ij}\xi_j}} \nonumber\\
 &= \int \mathcal D[\phi,\bar \phi,\bar h] e^{\int_{\bx} \{j_i \phi_i + i \bar j_i \bar \phi_i + i \bar k_i \bar h_i\}} e^{ - \mathcal{S}}\nonumber\\
 \text{with } \mathcal{S} &= i \int_{\bx} \Big\{ \bar \phi_i(\bx) \big[\p_t \phi_i(\bx) - F_i(\bx)\big] + \bar h_i H_i \Big\} + \frac{1}{2} \int_{\bx,\bx'} (\bar \phi_i G_{ij})(\bx) D_{jk}(\bx-\bx')(G_{k\ell}\bar \phi_\ell)(\bx')\,,
\end{align}
 using the property (\ref{eq:xi}) to compute the average value in the second line. Finally, one usually absorbs the complex $i$ in a redefinition of the response fields.

In the literature concerned with field theories for turbulence, it is more customary to integrate out the incompressibility constraint. This  allows one to get rid of the pressure fields at the price of having to deal with non-local transverse projector for the
velocity fields. Here, we choose to keep the pressure in the action. Although it doubles the number of fields, 
   one can easily show that the pressure sector of the action is not renormalized and keeps its bare form \cite{Canet16}.
  As a consequence, it is very simple to handle. The final action for Navier Stokes reads
\begin{align}
 {\cal S}[\vv,\bar{\vv},p,\bar p] &= \int_{\bx}\Big\{\bar v_\alpha(\bx)\Big[ \p_t 
v_\alpha(\bx) 
 -\nu \nabla^2 v_\alpha(\bx) +v_\beta (\bx)\p_\beta v_\alpha(\bx)+\frac 1\rho \p_\alpha p (\bx) \Big] + \bar p (\bx)\,\p_\alpha v_\alpha(\bx)\Big\}\nonumber\\
 &-\int_{t,\vec x,\vec x'}\bar 
v_\alpha(t,\vec x) N_{L^{\text{-}1}}(|\vec x-\vec x'|)\bar v_\alpha(t,\vec  x')\, .
\end{align}

\section{Renormalisation group flow equation and fixed point solution}
\label{AP:RG}

In this work, we focus on fully developed turbulence, that is we assume a large separation between the Kolmogorov scale
related to the dissipation and the integral scale related to the forcing. We probe the system at wave-vectors in the
inertial range, that is far from both above-mentioned scales. As previously mentioned, turbulence is phenomenologically
 known to  exhibit some form of scale invariance in this regime. 
This  suggests to use  \RG techniques to study turbulence, and indeed, these techniques were first applied in this context 
 back in the seventies \cite{Forster77,Dominicis79}. 
In this appendix, we explain  the link between scale invariance
 and fixed-point solutions of the \RG flow equation.
For this, we follow the presentation of~\cite{Delamotte16}, and concentrate on the generating functional of cumulant, ${\cal W}_{\kappa}$. 
For a generic out of equilibrium field theory with regulator, the scale-dependent generating functional is
\begin{equation}
 {\cal Z}_{\kappa}[j] = e^{{\cal W}_\kappa[J]} = \int D[\phi] e^{- {\cal S} - \Delta {\cal S}_\kappa + \int_{\bx} j \cdot \phi }\,.
\label{eq:Zgen}
\end{equation}
where $\phi$ is the $2n-$component combined field and response of the theory,
 \ie $\phi = (\phi_1, \dots \phi_n, \bar \phi_1, \dots \bar \phi_n)$, and $j$ is the corresponding $2n-$component
source. In the field theory of Navier-Stokes, the fields and response fields are $\vv, p$ and $\bar \vv, \bar p$ respectively. 
The regulator term $\Delta {\cal S}_\kappa$ is  quadratic in the fields,
 and depends explicitly on the scale $\kappa$, thus breaking scale invariance. Without this term and for a critical 
  theory (that is a theory whose parameters are at a critical point), the 
generalized correlation functions of the theory are expected to exhibit scale invariance, at least within a certain range of 
scales, which means that they are expected to be invariant under the following change of variables:
\begin{equation}
 \phi_i(t,\vx) \to b^{h_{i}} \phi_{i}(b^z t, b \vx)\,,
\end{equation}
with $h_i,\,z$ the scaling exponents of the corresponding critical theory. To make this statement more formal,
let us perform this change of variable inside ${\cal Z}_{\kappa}[j]$, and take the infinitesimal 
limit $b = e^{\epsilon} = 1 + \epsilon + o(\epsilon)$. One obtains,
\begin{align}
 \phi_i(t,\vx) &\to \phi_i(t,\vx) + \epsilon\, \delta \phi_i(t,\vx) + o(\epsilon)\,,\nonumber\\
 \text{ with }\quad \delta \phi_i(t,\vx) &= (h_i + \vx\cdot\p_{\vx} + z t\p_t)\phi_i(t,\vx)\,,
\end{align}
and  the corresponding variations of the action and of the regulator term follow as
\begin{align}
 {\cal S} &\to {\cal S} + \epsilon \,\delta S + o(\epsilon)\,,\nonumber\\
 \Delta {\cal S}_\kappa &\to \Delta {\cal S}_\kappa + \epsilon \delta \Delta {\cal S}_\kappa + o(\epsilon)\,.
\end{align}
The jacobian
of the change of variable is non-zero but it is constant in the fields so it does not play a role in the correlation functions
and it is omitted in the following. Identifying the terms of order $\epsilon$, one obtains the following equality:
\begin{equation}
 \label{eq:WIdilat1}
 \mean{\delta {\cal S}} + \mean{\delta \Delta {\cal S}_\kappa} = \int_{\bx} j\cdot \delta \mean{ \phi} \,.
\end{equation}
The form of $\Delta {\cal S}_\kappa$ can be chosen such that $\delta \Delta {\cal S}_\kappa = - \frac{d}{ds} \Delta {\cal S}_\kappa$, where $s$ is the ``\RG-time'' $s \equiv \ln{(\kappa/\Lambda)}$~\cite{Ellwanger94}.
Indeed, the generic form of $\Delta {\cal S}_\kappa$ is
\begin{equation}
 \Delta {\cal S}_\kappa = \frac{1}{2} \int_{\bx,\bx'} {\cal R}^{ij}_\kappa(\bx-\bx') \phi_i(\bx)\phi_j(\bx')\,,
\end{equation}
and one can choose ${\cal R}^{ij}_\kappa(\bx-\bx') = R^{ij}_\kappa\, \hat r_{ij}\big(\kappa(\vx-\vx'),\kappa^z(t-t')\big)$
 with $R^{ij}_\kappa \sim \kappa^{2d+2z-h_i-h_j}$, to enforce the
previous property.
This means that, for this form of regulator, the variation due to dilatation of space-time and fields is equal to the 
variation due to a dilatation of the renormalization scale.
For \NS, the regulators are $R_\kappa$ and $N_\kappa$ defined
in equation (\ref{eq:deltaSk}). For this property to hold, we require the following form
for their Fourier transform:
 \begin{equation} 
 N_\kappa(\vp) \equiv D_\kappa \hat n({\vp/\kappa}) \quad\quad\quad \hbox{and}\quad\quad\quad  R_\kappa(\vp) \equiv \nu_\kappa \hat r(\vp/\kappa) \, . 
 \end{equation}
with $D_\kappa \sim \kappa^{d+z-2h_{\bar v}}$ and $\nu_\kappa \sim \kappa^{d+z-h_v-h_{\bar v}}$. 

In the
general setting,  taking a $s$ derivative of \Eq{eq:Zgen}  entails
 the identity
  $\p_s \mean{\Delta {\cal S}_\kappa} = - \p_s {\cal W}_\kappa$~\footnote{We use a partial derivative to indicate that the sources
are kept fixed.}. Furthermore, the averages
of the fields can be expressed as the first functional derivatives of ${\cal W}_\kappa$, such that (\ref{eq:WIdilat1}) can  be written as
\begin{equation}
\p_s {\cal W}_\kappa + \mean{\delta {\cal S}} = \int_{\bx} j_i (h_i + \vx\cdot\p_{\vx} + z t\p_t)\frac{\delta {\cal W}_\kappa}{\delta j_i(\bx)}\,.
\label{eq:WIdilat2}
\end{equation}
The \rhs of the above equation is the variation of ${\cal W}_\kappa$ in a dilatation and the \lhs is the breaking of the scale
invariance, coming from the regulator and from the microscopic action respectively. Even if the theory is invariant under 
dilatations  (\ie if $\delta {\cal S} = 0$), the presence of the renormalization scale $\kappa$  breaks
scale invariance. To see that a fixed point of the \RG flow corresponds to scale invariance, 
 one  has to choose $\kappa$ as the unit of scale for space-time and the fields, and to introduce dimensionless
  quantities.
Defining
\begin{equation}
 {\cal W}_\kappa[j] = \hat {\cal W}_\kappa[\hat j]\quad \text{, with }\quad \hat j_i(\hat t, \hat \vx) = \kappa^{h_i-d-z} j_i(\hat x/\kappa, \hat t/\kappa^z) \,, 
\end{equation}
and replacing in (\ref{eq:WIdilat2}), one obtains that
\begin{equation}
\p_s \hat {\cal W}_\kappa = -\mean{\delta {\cal S}} \,.
\label{eq:WIdilat3}
\end{equation}
This equation shows that asking for a fixed point of $\hat {\cal W}_\kappa$ is equivalent to asking for a critical phenomenon. For critical phenomena, the scale invariance breaking term on the \lhs becomes sub-leading when 
$\kappa$ is below the momentum scale at which the system is probed.

Within the \NPRG framework, $\Gamma_\kappa$ is analytic, so that the singularities of ${\cal W}_\kappa$, the Legendre transform of 
$\Gamma_\kappa$, are  well controlled. Here, because of the regulators in the velocity and response velocity sector, the only infrared singularity comes from the pressure sector and gives the transverse projectors in the correlation functions.
 This allows one 
   to approximate ${\cal W}_\kappa$ for $\kappa \neq 0$, contrary to the original generating functional ${\cal W}$ without regulator, using functional Taylor expansions.
 The exact \RG flow of ${\cal W}_\kappa$ is given by (\ref{eq:dkw}).
This equation reads for the dimensionless generating functional $\hat {\cal W}_\kappa$
\begin{equation}
 \int_{\hat \bx} \hat j_i (h_i + \hat \vx\cdot\p_{\hat \vx} + z \hat t\p_{\hat t})\frac{\delta \hat {\cal W}_\kappa}{\delta \hat j_i(\hat \bx)} = -\p_s \hat {\cal W}_\kappa - \frac{1}{2}\, \int_{\hat \bx,\hat \by}\! (2d+2z-h_i-h_j + \hat \vx \cdot \p_{\hat \vx} + z \hat t \p_{\hat t})\hat r_{ij}(\hat \bx - \hat \by) \,\Big\{\frac{\delta^2 \hat {\cal W}_\kappa}{\delta \hat j_{i}({\hat \bx}) \delta \hat j_{j}({\hat \by})} + \frac{\delta \hat {\cal W}_\kappa}{\delta \hat j_{i}({\hat \bx})} \frac{\delta \hat {\cal W}_\kappa}{\delta \hat  j_{j}({\hat \by})}\Big\}
\end{equation}
In standard critical phenomena, both terms of the \rhs are sub-leading compared to the terms in the \lhs.
 On the one hand, the first term
is sub-leading by definition of critical phenomena through (\ref{eq:WIdilat3}). On the other hand,
 the second term is sub-leading if the \RG flow equation
decouples the fast variables of the system from the slow ones. This decoupling property
 is satisfied by the \RG flows corresponding to equilibrium phase transitions for example, 
 and also to many critical phenomena out-of-equilibrium \cite{Canet04b,Canet05,Canet10}.
  As a consequence, the correlation functions of these systems exhibit scale invariance.
 However, in the field theory of Navier-Stokes, we show in this paper that while the first term is sub-leading
  (that is, there is a fixed-point), 
the second term is not. This peculiar fact entails a breaking of scale invariance for the correlation functions of turbulence.

\section{Ward identity for extended Galilean invariance}
\label{AP:Ward}

In this appendix, we derive the general Ward identity associated with the extended Galilean symmetry for an arbitrary vertex function $\Gamma^{(m,n)}$. We consider the functional Ward identity \eq{eq:wardGal} derived in \citep{Canet15}, 
\begin{equation}
 \int_{\vx} \Big\{\big(\delta_{\alpha\beta}\p_t +  \p_\beta 
u_\alpha\big)\frac{\delta \Gamma_\kappa}{\delta u_\alpha}   \nonumber
+ \p_\beta \bar u_\alpha \frac{\delta \Gamma_\kappa}{\delta \bar 
u_\alpha} \Big\}= -\int_{\vx} \p_t^2 \bar u_\beta \, ,
\end{equation}
where  the pressure terms are omitted since 
 they give no contribution in the following derivation.
Taking $m$ functional derivatives of this identity with respect to velocity fields $u_{\alpha_i}(\bx_i)$ -- $i=1,\dots,m$ -- and $n$ with respect to response velocity fields $\bar u_{\alpha_{j}}(\bx_{j})$ -- $j=m,\dots,m+n$ --, and setting the fields to zero yields
\begin{equation}
  \int_{\vx} \Big\{ \p_t \Gamma_{\alpha\alpha_1\dots\alpha_{m+n}}^{(m+1,n)}(\bx, \bx_1, \bx_{m+n}) - \sum_{k=1}^{m+n} \delta(t-t_k)\delta^d(\vx-\vx_k)\p_\alpha \Gamma_{\alpha_1\dots\alpha_{m+n}}^{(m,n)}( \overbrace{\bx_i}^{k-1},\,\bx,\overbrace{\bx_j}^{m+n-k}) \Big\} =0 \, .
\end{equation}
This identity can be expressed in  Fourier space. It yields in terms of the Fourier transforms $\tilde \Gamma^{(k,\ell)}$ 
\begin{equation}
  \tilde \Gamma^{(m+1,n)}_{\alpha\alpha_1\dots\alpha_{n+m}}(\omega,\vp=0,\bp_1, \dots, \bp_{n+m})=
 -\sum_{k=1}^{n+m}\frac{p_k^{\alpha}}{\omega}\tilde \Gamma^{(m,n)}_{\alpha_1\dots\alpha_{n+m}}(\overbrace{\bp_i}^{k-1},\,\omega_k+\omega,\vp_k,\overbrace{\bp_j}^{m+n-k}) \, .
\end{equation}
The identity for the reduced Fourier transform $\bar \Gamma^{(m+1,n)}$  straightforwardly follows from it. One obtains:
\begin{equation}
  \bar \Gamma^{(m+1,n)}_{\alpha\alpha_1\dots\alpha_{n+m}}(\omega,\vp=0,\bp_1, \dots, \bp_{n+m-1})
  = -\sum_{k=1}^{n+m-1}\frac{p_k^{\alpha}}{\omega}
 \Bigg[\bar \Gamma^{(m,n)}_{\alpha_1\dots\alpha_{n+m}}(\overbrace{\bp_i}^{k-1},\,\omega_k+\omega,\vp_k,\overbrace{\bp_j}^{m+n-1-k}) - \bar \Gamma^{(m,n)}_{\alpha_1\dots\alpha_{n+m}}(\overbrace{\bp_i}^{m+n-1}) \Bigg] \, .
\end{equation}

\section{Derivation of a closed flow equation for any $n$-point correlation function}
\label{AP:flowN}

\subsection{Flow equation for a generic $n$-point function}

Let us derive the flow equation for a generic  generalized $n$-point connected correlation function $\bar G^{(n)}$.
 It is obtained by taking $n$ functional derivatives of \eq{eq:dkw} with respect to the sources $j_{\alpha_k}$, $k=1,\dots,n$,
 which yields
\begin{align}
 \p_\kappa G^{(n)}_{\alpha_1\dots\alpha_{n}}[{\bx_1}, \dots, {\bx_n};j] &= - \frac{1}{2}\,  \int_{\by_1,\by_2}\! \p_\kappa [{\cal R}_\kappa]_{ij}({\by_1-\by_2}) \,\Bigg\{ G^{(n+2)}_{ij \alpha_1\dots\alpha_{n}}[\by_1,\by_2,\bx_1,\dots,\bx_n;j]
\nonumber\\
 & + \sum_{(\{\alpha_k\}, \{\alpha_\ell\}) \atop k+\ell=n} G^{(k+1)}_{i \{\alpha_k\}}[{\by_1,\{\bx_{k}\}};j] G^{(\ell+1)}_{j \{\alpha_\ell\}}  [{\by_2,\{\bx_{\ell}\}};j]\Bigg\} \, .
\label{eq:A1}
\end{align}
In this equation, the indices stand for both the type of source  $\vJ$ or $\bar \vJ$ and the space component,
 and $(\{\alpha_k\}, \{\alpha_\ell\})$ indicates all the possible bipartitions of the $n$ indices $\alpha_i$, and $(\{\bx_{k}\},\{\bx_{\ell}\})$ the corresponding bipartition in coordinates. 
Let us concentrate  on the first line of \eq{eq:A1}. One can write
\begin{align}
 &\int_{\by_1,\by_2}\! \p_\kappa [{\cal R}_\kappa]_{ij} ({\by_1-\by_2}) \, G^{(n+2)}_{ij \alpha_1\dots\alpha_{n}}[\by_1,\by_2,\bx_1,\dots,\bx_n;j] = \int_{\by_1,\by_2}\! \p_\kappa [{\cal R}_\kappa]_{ij}({\by_1-\by_2}) \nonumber\\
 & \times \Bigg[\int_{\bz_1,\bz_2} G^{(2)}_{k i}[\bz_1,\by_1;j]  G^{(2)}_{\ell j}[\bz_2,\by_2;j]
  \frac{\delta^2}{\delta \varphi_k(\bz_1)\delta\varphi_\ell(\bz_2)} + \int_{\bz} G^{(3)}_{\ell ij}[\bz,\by_1,\by_2;j]  \frac{\delta}{ \delta\varphi_\ell(\bz)} \Bigg]
 G^{(n)}_{\alpha_1\dots\alpha_{n}}[{\bx_1}, \dots, {\bx_n};j] \, .
\label{eq:A2}
\end{align}
The derivatives of $G^{(n)}$ with respect to $\varphi$ must be understood as acting on  $G^{(n)}$ viewed as a diagram constructed from $\Gamma_{\kappa}$ vertices. More precisely, $G^{(n)}$ is the sum of all tree diagrams with vertices the $\Gamma^{(k)},k\leq n$ and with edges the propagator $G^{(2)}$, the latter satisfying
\begin{equation}
 G^{(2)}_{k\ell}[\bx,\by;j] = \frac{\delta \varphi_k(\bx)}{\delta j_\ell(\by)} = \left(\frac{\delta j}{\delta \varphi}\right)^{-1}_{k\ell}(\bx,\by) = [\Gamma^{(2)} + {\cal R_{\kappa}}]^{-1}_{k\ell}[\bx,\by;\varphi],
 \label{eq:g2gam2}
\end{equation}
using the property of the Legendre transform \eq{eq:legendre}.
Furthermore, introducing the differential operator
\begin{equation}
 \tilde \p_\kappa \equiv \p_\kappa R_\kappa \frac{\delta}{\delta R_\kappa} + \p_\kappa N_\kappa \frac{\delta}{\delta N_\kappa} \, ,
\end{equation}
and using the expression \eq{eq:g2gam2}, one has
 \begin{equation}
 \tilde \p_\kappa G^{(2)}_{k \ell}[\bz_1,\bz_2;j] = -\int_{\by_1,\by_2} 
  \p_\kappa [{\cal R}_\kappa]_{ij}({\by_1-\by_2})  G^{(2)}_{k i}[\bz_1,\by_1;j]  G^{(2)}_{\ell j}[\bz_2,\by_2;j] \, ,
\label{eq:flowdtildeC}
\end{equation}
which appears in the first term of the \rhs of \eq{eq:A2}.
On the other hand,  the second term  in the \rhs of \eq{eq:A2} vanishes when the fields are set to zero, since it is proportional to the flow of the average velocity.
Indeed, the functions $G^{(1)}_{i}(\bx)$ are the expectation values of the velocity fields and response velocities, which are zero at zero fields, and so are their flow equations. The expression of their flow can be deduced by 
taking one derivative of \eq{eq:dkw} with respect to a field and setting the fields to zero, which yields
\begin{equation}
 \p_\kappa G^{(1)}_\ell(\bz) = - \frac 1 2 \int_{\by_1,\by_2}  \p_\kappa [{\cal R}_\kappa]_{ij}({\by_1-\by_2}) 
\, ,
\end{equation}
omitting additional contribution proportional to  $G^{(1)}$. By identification,
 one concludes that the second term in the \rhs of \Eq{eq:A2} vanishes when evaluated at zero fields.
Gathering  the previous expressions and setting the fields to zero, the flow equation for $G^{(n)}$ may be rewritten as
\begin{align}
 \p_\kappa  G^{(n)}_{\alpha_1\dots\alpha_{n}}({\bx_1}, \dots, {\bx_n}) =& \frac{1}{2}\, \int_{\by_1,\by_2}\! \Bigg\{ 
\tilde \p_\kappa G^{(2)}_{k \ell}(\by_1,\by_2) \left[\frac{\delta^2}{\delta \varphi_k(\by_1)\delta\varphi_\ell(\by_2)} 
G^{(n)}_{\alpha_1\dots\alpha_{n}}[{\bx_1}, \dots, {\bx_n};j]\right]_{\varphi=0}\nonumber\\
 &+  \sum_{(\{\alpha_k\}, \{\alpha_\ell\}) \atop k+\ell=n} G^{(k+1)}_{i \{\alpha_k\}}({\by_1,\{\bx_{k}\}})\p_\kappa [{\cal R}_\kappa]_{ij}({\by_1-\by_2}) G^{(\ell+1)}_{j \{\alpha_\ell\}}  ({\by_2,\{\bx_{\ell}\}}) \,\Bigg\} \, .
\end{align}
This yields in Fourier space
\begin{align}
 \p_\kappa \bar G^{(n)}_{\alpha_1\dots\alpha_{n}}({\bp_1}, \dots, {\bp_{n-1}}) =& \frac{1}{2}\,  \int_{\bq}\!
\tilde \p_\kappa \bar G^{(2)}_{k \ell}(\bq) \left[\frac{\delta^2}{\delta \varphi_k(\bq)\delta\varphi_\ell(-\bq)}
  \bar G^{(n)}_{\alpha_1\dots\alpha_{n}}[{\bp_1}, \dots, {\bp_{n-1}};j]\right]_{\varphi=0}\nonumber\\
 & -  \frac{1}{2}\,  \sum_{(\{\alpha_k\}, \{\alpha_\ell\}) \atop k+\ell=n} \bar G^{(k+1)}_{\{\alpha_k\} i}(\{\bp_{k}\})\p_\kappa [{\cal R}_\kappa]_{ij}(\sum \{\bp_{k}\}) \bar G^{(\ell+1)}_{\{\alpha_\ell\} j}  (\{\bp_{\ell}\}) \, ,
\label{eq:A3}
\end{align}
with the definition $\sum \{\bp_{k}\} = \sum_{i=1}^k \bp_{i}$ and $\sum \{\bp_{k}\} + \sum \{\bp_{\ell}\} =0$ and 
where in the first line the Fourier transform is meant after the functional derivatives
\begin{equation}
 \left[\frac{\delta^2}{\delta \varphi_k(\bq)\delta\varphi_\ell(-\bq)}
  \bar G^{(n)}_{\alpha_1\dots\alpha_{n}}[{\bp_1}, \dots, {\bp_{n-1}};j]\right]_{\varphi=0} \equiv \text{FT}\Bigg(\frac{\delta^2}{\delta \varphi_k(\bz_1)\delta\varphi_\ell(\bz_2)} G^{(n)}_{\alpha_1\dots\alpha_{n}}[{\bx_1}, \dots, {\bx_n};j]\Big|_{\varphi=0}\Bigg)(\bq,-\bq, \bp_1,\dots\bp_{n-1}) \,
 \label{eq:symbolic}
\end{equation}
with $\text{FT}(\dots)$ denoting the Fourier transform.

 Let us now show thatthis flow equation can be closed at leading order at large wave-numbers, \ie expressed in terms of $\bar G^{(k)}$ with $k\le n$ only, using the Ward identities associated with the extended Galilean and extended shift symmetries.

\subsection{Limit of large wave-numbers}

We focus on the flow equation \eq{eq:A3}, and now consider the limit of large wave-numbers, which we define as all external wave-numbers  being large compared to the \RG scale $|\vp_i| \gg \kappa$, as well as all possible  partial sums being large  $\sum \{\bp_{k}\} \gg \kappa$, which means that we exclude  exceptional configurations where a partial sum vanishes. The following proof relies on the presence of the (derivative of the) regulator term $\p_\kappa [{\cal R}_\kappa]$ in the flow equation \eq{eq:A3}. The key properties of this term are that, on the one hand, it rapidly  tends to zero for wave-numbers greater that the \RG scale, \Eq{eq:limitreg}, and on the other hand, it ensures the analyticity of all vertex functions at any finite $\kappa$. 
 
Let us examine the two terms in the  \rhs of \eq{eq:A3} in this limit. The second term is proportional to the regulator evaluated at a sum of external wave-numbers  $\sum \{\bp_{k}\} \gg \kappa$. Hence this term vanishes in the limit of large wave-numbers and all the corresponding graphs are negligible. Thus  only the first term survives in this limit.

For convenience, we now switch to the $\tilde G^{(n)}$ Fourier transforms, instead of the $\bar G^{(n)}$. The flow equation of $\tilde G^{(n)}$ reads 
\begin{equation}
 \p_\kappa \tilde G^{(n)}_{\alpha_1\dots\alpha_{n}}({\bp_1}, \dots, {\bp_{n}}) =  \frac{1}{2}\,  \int_{\bq_1,\bq_2}\!
 \tilde \p_\kappa \tilde  G^{(2)}_{k \ell}(-\bq_1,-\bq_2)\left[ \frac{\delta^2}{\delta \varphi_k(\bq_1)\delta\varphi_\ell(\bq_2)} 
 \tilde G^{(n)}_{\alpha_1\dots\alpha_{n}}[{\bp_1}, \dots, {\bp_{n}};j] \right]_{\varphi=0}\, , \label{eq:flowGn}
\end{equation}
with the same  notation as in \Eq{eq:symbolic}.
Let us first show that only the derivatives with respect to the velocity fields, \ie $\varphi_k=\varphi_\ell=u_\alpha$, 
contribute in the limit of large wave-numbers.

\subsubsection{Action of derivatives with respect to response velocity fields}
 
 Any connected correlation function $G^{(n)}$  can be expressed in terms of $\Gamma^{(m)}$ vertex functions,
  $m=3,\dots,n$ and propagators $G^{(2)}$ as a sum of tree diagrams. This means that  a $\varphi_\ell(\bq)$ derivative,  with
 $\bq\equiv(\varpi,\vq)$, 
  either acts on a vertex functions or on a propagator. Let us consider the action of a response
   field $\bar u_\mu(\bq)$  on a vertex function $\Gamma^{(k,\ell)}$ with $k+\ell=m$:
\begin{equation}
\frac{\delta}{\delta \bar u_\mu(\bq)}\tilde \Gamma_{\alpha_1 \dots \alpha_{k+\ell}}^{(k,\ell)}[\bp_1,\dots,\bp_{k+\ell};\varphi] =  \tilde \Gamma_{\alpha_1\dots\alpha_k \mu \dots \alpha_{k+\ell}}^{(k,\ell+1)}[\bp_1,\dots,\bp_k,\varpi,\vq,\dots,\bp_{k+\ell};\varphi] \, ,
 \end{equation} 
with  the same convention as in \eq{eq:flowGn}.
 The wave-number $|\vq|$ carried by the $\bar u_{\mu}$ field is cut to values $|\vq| \lesssim \kappa$ by the derivative of the regulator in the flow equation \eq{eq:flowGn}. Hence it is negligible compared  to the modulus of all external wave vectors $\vp_i$ and all their possible partial sums, and can be safely set to zero in \PI vertex functions. This vertex function then vanishes due to the Ward identity related to the time-gauged shift symmetry \Eq{eq:wardShiftN}. 
One deduces that in the limit of large wave-numbers
\begin{equation}
\frac{\delta}{\delta \bar u_\mu(\bq)}\tilde \Gamma_{\alpha_1\dots\alpha_{k+\ell}}^{(k,\ell)}[\bp_1,\dots,\bp_{k+\ell};
\varphi]\Big|_{\varphi=0} \underset{|\vp_i|\gg \kappa}{=}  
\tilde \Gamma_{\alpha_1\dots\alpha_k \mu\dots\alpha_{k+\ell}}^{(k,\ell+1)}(\bp_1,\dots,\bp_k,\varpi,\vec 0,\dots,
\bp_{k+\ell}) \, =0 \, .\label{eq:A5bis}
 \end{equation}
We now consider the action of the response field $\bar u_\mu(\bq)$ on a propagator:
\begin{align}
 \frac{\delta}{\delta \bar u_\mu(\bq)}
 \tilde G^{(2)}_{\alpha_1 \alpha_2}[\bp_1,\bp_2;j]\Big|_{\varphi=0}  &= - \int_{ \bk_1,\bk_2}
 \tilde G^{(2)}_{\alpha_1 i}(\bp_1,-\bk_1) \tilde \Gamma^{(3)}_{i j \mu} (\bk_1,\bk_2,\bq)\tilde G^{(2)}_{j\alpha_2}(-\bk_2,\bp_2)  \nonumber\\
  &\hspace{-0.3cm}\underset{|\vp_i|\gg \kappa}{=}  \int_{\bk_1,\bk_2} \tilde G^{(2,0)}_{\alpha_1 i}(\bp_1,-\bk_1)\big[  i 
 k_1^{i} \delta_{\mu j} + i k_2^j \delta_{\mu i}   \big])\tilde G^{(2,0)}_{j\alpha_2}(-\bk_2,\bp_2) \nonumber\\
 &=0 \, .\label{eq:A5}
\end{align} 
Indeed, in the first line, there are three non-zero contributions, one proportional to $\tilde \Gamma^{(1,2)}$, one to $\tilde \Gamma^{(0,3)}$, and one to $\tilde \Gamma^{(2,1)}$, with the $\vq$ wave-vector carried by a response field. In the limit of large wave-numbers, one can set again $\vq=0$, and both vertex functions are then fixed by a shift Ward identity. The identity \eq{eq:wardShiftN} applies for 
 $\tilde \Gamma^{(1,2)}$ and $\tilde \Gamma^{(0,3)}$ so these contributions vanish, and the identity \eq{eq:wardShift21} applies for $\tilde \Gamma^{(2,1)}$, which corresponds to the second equality in \Eq{eq:A5}. Then this contribution vanishes because of incompressibility (which implies that $G^{(2)}$ is transverse, so that the contraction with the  wave-vectors $\vk_1$ and $\vk_2$ is zero).

Let us now consider the action of the two derivatives in \eq{eq:flowGn} where at least one is a response velocity, where
 again $\tilde G^{(n)}$ is considered as a sum of tree diagrams constructed with vertices $\Gamma^{(m)}$ and propagators. If
  each of the derivatives $\delta/\delta \bar u_{\mu_1}(\bq_1)$ and $\delta/\delta \varphi_{\mu_2}(\bq_2)$ acts
   on a different element of the diagram, then, when evaluated at zero fields, 
  the corresponding term vanishes because of \Eqs{eq:A5bis} or \eq{eq:A5}. Let us now examine the contributions 
  where the two derivatives act on
   the same element. If this element is a vertex function, then this term also vanishes because 
  of the shift symmetry at large wave-numbers
 \begin{equation}
\frac{\delta^2}{\delta \bar u_{\mu_1}(\bq_1)\delta u_{\mu_2}(\bq_2)} \tilde \Gamma_{\alpha_1\dots\alpha_{k+\ell}}^{(k,\ell)}[\bp_1,\dots,\bp_{k+\ell};
\varphi]\Big|_{\varphi=0} \underset{|\vp_i|\gg \kappa} = \tilde \Gamma_{{\mu_2}\alpha_1\dots\alpha_k \mu_1\dots\alpha_{k+\ell}}^{(k+1,\ell+1)}(\varpi_2,\vec 0,\bp_1,\dots,\bp_k,\varpi_1,\vec 0,\dots,
\bp_{k+\ell}) \, =0 \, ,
 \end{equation} 
and similarly if $\varphi$ is a response velocity. Finally, the same result holds if the two derivatives act on a propagator
\begin{align}
 &\frac{\delta^2}{\delta \bar u_{\mu_1}(\bq_1)\delta \varphi_{\mu_2}(\bq_2)} 
 \tilde G^{(2)}_{\alpha_1 \alpha_2}[\bp_1,\bp_2;j]\Big|_{\varphi=0} = - \int_{ \bk_1,\bk_2} \tilde G^{(2)}_{\alpha_1 i}(\bp_1,-\bk_1) \tilde \Gamma^{(4)}_{i j \mu_1{\mu_2}} (\bk_1,\bk_2,\bq_1,\bq_2)
 \tilde G^{(2)}_{j\alpha_2}(-\bk_2,\bp_2)  \nonumber\\
 &\quad\quad \quad+ \int_{ \bk_i} \tilde G^{(2)}_{\alpha_1 i}(\bp_1,-\bk_1) \tilde \Gamma^{(3)}_{i j \mu_1} (\bk_1,\bk_2,\bq_1)
 \tilde G^{(2)}_{j k}(-\bk_2,-\bk_3) \tilde \Gamma^{(3)}_{k\ell {\mu_2}} (\bk_3,\bk_4,\bq_2)
  \tilde G^{(2)}_{\ell\alpha_2}(-\bk_4,\bp_2) + (\mu_1,\bq_1) \longleftrightarrow  (\mu_2,\bq_2) \nonumber\\
  &\hspace{5.5cm}\underset{|\vp_i|\gg \kappa}{=}  0 \, . 
\end{align} 
where the permutation $1\leftrightarrow 2$ applies only for the second line. The last identity is zero because the vertices $\tilde \Gamma_{i j \mu_1{\mu_2}}^{(4)}$
 and $\tilde \Gamma^{(3)}_{i j \mu_1}$ vanish at large wave-numbers when setting $\vq_1$ to zero 
  because of the shift symmetry and incompressibility as in \Eq{eq:A5}.
One concludes that if at least one of the derivatives in \eq{eq:flowGn} is a response velocity,
 the corresponding term vanishes. Thus, only the velocity fields $\varphi_k(\bq)=u_\mu(\bq)$ contribute in \Eq{eq:flowGn}, that is
\begin{equation}
 \p_\kappa \tilde G^{(n)}_{\alpha_1\dots\alpha_{n}}({\bp_1}, \dots, {\bp_{n}}) = \frac{1}{2}\,  \int_{\bq_1,\bq_2}\! 
\tilde \p_\kappa \tilde G^{(2,0)}_{\mu_1 \mu_2}(-\bq_1,-\bq_2)\left[ \frac{\delta^2}{\delta u_{\mu_1}(\bq_1)\delta u_{\mu_2}(\bq_2)}  \tilde G^{(n)}_{\alpha_1\dots\alpha_{n}}[{\bp_1}, \dots, {\bp_{n}};j] \right]_{\varphi=0}\, .
\end{equation}

\subsubsection{Action of one derivative with respect to a velocity field}

The aim is now to show that this flow equation can be simply expressed in the limit of large wave-numbers in term of the finite difference  operator ${\cal D}_{\mu}(\varpi)$ defined in \Eq{eq:wardGalN}. In fact, we  establish below the following property: \\

 $(P)$: {\it taking a functional derivative with respect to $u_\mu(\bq)$  with $\vq=0$ is equivalent at zero fields  to applying the operator ${\cal D}_\mu(\varpi)$.}\\

\noindent In this section, we adopt the following notation for the frequencies:
 $\bp_i \equiv(\vp_i,\omega_i)$, $\bq_i \equiv(\vq_i,\varpi_i)$ and $\bk_i \equiv(\vk_i,\nu_i)$, that is $\omega_i$ is an external frequency, $\varpi$ is the frequency associated with a regulated wave-vector, and $\nu_i$ is used for other internal frequencies. 
We first show that the property $(P)$ is true for $\tilde G^{(2)}$. One has 	
\begin{align}
 \frac{\delta}{\delta u_\mu(\bq)} &\tilde G^{(2)}_{\alpha_1 \alpha_2}[\bp_1,\bp_2;j]\Big|_{\varphi=0} \nonumber\\
 &=  - \int_{\bk_1,\bk_2} \tilde G^{(2)}_{\alpha_1 i}(\bp_1,-\bk_1) \Gamma^{(3)}_{\mu i j } (\bq,\bk_1,\bk_2)\tilde G^{(2)}_{j\alpha_2}(-\bk_2,\bp_2)  \nonumber\\
 &\hspace{-0.3cm}\underset{|\vp_i|\gg \kappa}{=} \int_{\bk_1,\bk_2} \tilde G^{(2)}_{\alpha_1 i}(\bp_1,-\bk_1) \Big[\frac{k_1^{\mu}}{\varpi}\tilde\Gamma^{(2)}_{ i j } (\nu_1+\varpi,\vk_1,\bk_2) + \frac{k_2^{\mu}}{\varpi}\tilde\Gamma^{(2)}_{ i j } (\bk_1,\nu_2+\varpi,\vk_2) \Big]\tilde G^{(2)}_{j\alpha_2}(-\bk_2,\bp_2)  \nonumber\\
 &=  \int_{\bk_1}  \frac{k_1^{\mu}}{\varpi} \tilde G^{(2)}_{\alpha_1 i}(\bp_1,-\bk_1) \delta_{i\alpha_2}\delta(\varpi+\nu_1+\omega_2)\delta^d(\vk_1+\vp_2) +  \int_{\bk_2} \frac{k_2^{\mu}}{\varpi}  \delta_{j\alpha_1} \delta(\varpi+\omega_1+\nu_2)\delta^d(\vk_2+\vp_1) \tilde G^{(2)}_{j\alpha_2}(-\bk_1,\bp_2)  \nonumber\\
 &=-\frac{p_1^{\mu}}{\varpi}\tilde G^{(2)}_{\alpha_1 \alpha_2}(\omega_1+\varpi,\vp_1,\bp_2) - \frac{p_2^{\mu}}{\varpi}\tilde G^{(2)}_{ \alpha_1 \alpha_2}(\bp_1,\omega_2+\varpi, \vp_2) \nonumber\\
 &\equiv {\cal D}_\mu(\varpi)  \tilde G^{(2)}_{\alpha_1 \alpha_2}(\bp_1,\bp_2)\, .\label{eq:AG2}
 \end{align} 
In the second line, the limit of large wave-numbers is taken, such that the internal wave-vector $\vq$ can be set to zero in the \PI vertex functions $\tilde\Gamma^{(3)}$, and then  the Ward identity \eq{eq:wardGalN} associated with time-gauged Galilean invariance is used.\\

As already mentioned, any connected correlation function $G^{(n)}$  can be expressed as a sum of trees whose vertices are the
 $\Gamma^{(m)}$ vertex functions, $m=3,\dots,n$ and edges the propagators $G^{(2)}$.
   We will rely on this structure to prove $(P)$.
Let us first show that the property $(P)$ is verified for any ``star'' diagrams, which are depicted in Fig. 1. Let ${\cal S}$ be 
such a star diagram,
\begin{equation}
 \tilde {\cal S}^{(n)}_{\alpha_{1}\dots \alpha_n}(\bp_1,\dots,\bp_n) = \prod_{k=1}^n \int_{\{\bk_k\}} \tilde \Gamma^{(n)}_{i_{1}\dots i_n}(\bk_1,\dots,\bk_n) \tilde  G^{(2)}_{i_k \alpha_k}(-\bk_k,\bp_k) \, .
\end{equation}

  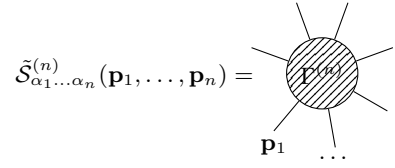
\begin{figure}
  \centering
  \begin{tikzpicture}
    \node[draw,circle, pattern=north east lines] (G) at (0,0) {$\Gamma^{(n)}$};
    \node[below] at (-130:1) {$\bp_1$};
    \node[left,scale=1] at (-180:0.8) {$\tilde {\cal S}^{(n)}_{\alpha_{1}\dots \alpha_n}(\bp_1,\dots,\bp_n) =$};
    \draw (-130:1) -- (G);
    \node[below] at (-80:1) {$\dots$};
    \draw (-80:1) -- (G);
    \draw[-] (-30:1) -- (G);
    \draw[-] (20:1) -- (G);
    \draw[-] (70:1) -- (G);
    \draw[-] (110:1) -- (G);
    \draw[-] (160:1) -- (G);
  \end{tikzpicture}
  \label{stardiag}
  \caption{Diagrammatic representation of a star diagram.}
 \end{figure}

If the derivative with respect to $u_\mu(\bq)$ acts on the vertex function, then in the limit of large wave-numbers, one has
 \begin{align}
 \frac{\delta}{\delta u_\mu(\bq)}\tilde \Gamma_{\alpha_1\dots\alpha_{k+\ell}}^{(k,\ell)}[\bp_1,\dots,\bp_{k+\ell};\varphi]\Big|_{\varphi=0} &=  \tilde \Gamma_{\mu\alpha_1\dots\alpha_{k+\ell}}^{(k+1,\ell)}(\varpi,\vq=0,\bp_1,\dots,\bp_{k+\ell})\nonumber\\
& =  -\sum_{j=1}^{k+\ell}\frac{p_j^{\mu}}{\varpi}\tilde \Gamma^{(k,\ell)}_{\alpha_1\dots\alpha_{k+\ell}}(\overbrace{\bp_i}^{j-1},\,\omega_j+\varpi,\vp_j,\overbrace{\bp_i}^{k+\ell-j}) \nonumber\\
&\equiv {\cal D}_\mu(\varpi)\tilde \Gamma_{\alpha_1\dots\alpha_{k+\ell}}^{(k,\ell)}(\bp_1,\dots,\bp_{k+\ell})\, . \nonumber\\
 \end{align} 
On the other hand, the action of the derivative $u_\mu(\bq)$ on a propagator is given by \Eq{eq:AG2}.
One hence obtains for the derivative of $\tilde {\cal S}^{(n)}$:
\begin{align}
 \frac{\delta}{\delta u_\mu(\bq)} &\tilde {\cal S}^{(n)}_{\alpha_{1}\dots \alpha_n}[\bp_1,\dots,\bp_n;j] \Big|_{\varphi=0} \nonumber\\
 &= \sum_{\ell=1}^n \prod_{{k=1} \atop {k\neq \ell}}^n \int_{\{\bk_k\}} \tilde  G^{(2)}_{i_k \alpha_k}(-\bk_k,\bp_k)\Big[\frac{k_\ell^\mu}{\varpi}  \tilde  G^{(2)}_{i_\ell \alpha_\ell}(-\nu_\ell +\varpi, -\vk_\ell,\bp_\ell) \nonumber\\
 &-\frac{p_\ell^\mu}{\varpi}  \tilde  G^{(2)}_{i_\ell \alpha_\ell}(-\bk_\ell,\omega_\ell +\varpi, \vp_\ell)\Big]  \tilde \Gamma^{(n)}_{i_{1}\dots i_n}(\bk_1,\dots,\bk_n) - \prod_{k=1}^n \int_{\{\bk_k\}} \tilde  G^{(2)}_{i_k \alpha_k}(-\bk_k,\bp_k)\sum_{\ell=1}^n \frac{k_\ell^\mu}{\varpi}  \tilde \Gamma^{(n)}_{i_{1}\dots i_n}(\dots,\nu_\ell+\varpi,\vk_\ell,\dots) \nonumber\\
 &= - \sum_{\ell=1}^n \,\frac{p^\mu_\ell}{\varpi} \, \prod_{{k=1} \atop {k\neq \ell}}^n \int_{\{\bk_k\}} \tilde  G^{(2)}_{i_k \alpha_k}(-\bk_k,\bp_k) \tilde G^{(2)}_{i_\ell \alpha_\ell}(-\bk_\ell,\omega_\ell+\varpi, \vp_\ell)\tilde \Gamma^{(n)}_{i_{1}\dots i_n}(\bk_1,\dots,\bk_n)\nonumber\\
& \equiv  {\cal D}_\mu(\varpi)\tilde {\cal S}^{(n)}_{\alpha_{1}\dots \alpha_n}(\bp_1,\dots,\bp_n) \, ,
\end{align}
noting that the two terms proportional to $k_\ell^\mu$ cancel out by wave-vector and frequency conservation. This
 can be checked  explicit by making the change of variable $\nu_\ell + \varpi \to \nu_\ell$ in the second term, which becomes equal to the
first term in square brackets. This cancellation is a consequence of the equality of the wave-vector and frequency exiting  $\bar G^{(2)}$ with the ones entering 
 the corresponding leg of  $\bar \Gamma^{(n)}$.
Thus, the property $(P)$ is true for any star diagrams, which is represented diagrammatically on Fig. 2.
 \begin{figure}
  \begin{tikzpicture}
    \node[draw,circle, pattern=north east lines] (G) at (0,0) {$\Gamma^{(n)}$};
    \node[below] at (-130:1) {$\bp_1$};
    \node[left,scale=1] at (-180:0.8) {$\displaystyle \frac{\delta}{\delta u_\mu(\bq)}$};
    \draw (-130:1) -- (G);
    \node[below] at (-80:1) {$\dots$};
    \draw (-80:1) -- (G);
    \draw[-] (-30:1) -- (G);
    \draw[-] (20:1) -- (G);
    \draw[-] (70:1) -- (G);
    \draw[-] (110:1) -- (G);
    \draw[-] (160:1) -- (G);
    \node[left, scale=1] at (3.3,0) {$\Bigg|_{\varphi=0}= \mathcal{D}_{\,\mu}(\varpi)$};
    \node[draw,circle, pattern=north east lines] (G1) at (4.2,0) {$\Gamma^{(n)}$};
    \draw  (G1) -- ++(-130:1);
    \draw  (G1) -- ++(-80:1);
    \draw[-] (G1) -- ++(-30:1);
    \draw[-] (G1) -- ++(20:1);
    \draw[-] (G1) -- ++(70:1);
    \draw[-] (G1) -- ++ (110:1);
    \draw[-] (G1) -- ++(160:1);
  \end{tikzpicture}
\caption{Diagrammatic representation of the property $(P)$ on a star diagram.}
 \end{figure}
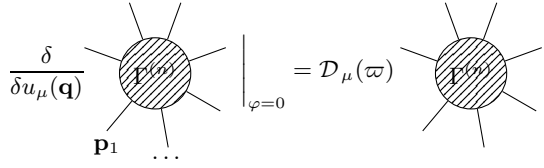

Let us now prove by induction that $(P)$ holds for any tree diagram. Let $n$ be a non-negative integer and let us assume that 
$(P)$ holds for all diagrams whose number of external legs is equal or lower than $n$.
Let us consider a diagram with $(n+1)$ external legs, denoted $\tilde {\cal T}^{(n+1)}$. If $\tilde {\cal T}^{(n+1)}$ is a star diagram, $(P)$ holds, otherwise
there exist two integers $m_1, m_2 \geq 2$ such that $m_1+m_2=n+1$ and two diagrams $\tilde {\cal T}^{(m_1+1)}$ and $\tilde {\cal T}^{(m_2+1)}$
such that
\begin{align}
 \tilde {\cal T}^{(n+1)}_{\alpha_1\dots \alpha_{m_1+m_2}}(\bp_1,\dots,\bp_{m_1+m_2}) 
 &= \int_{\{\bk_k\}} \tilde {\cal T}^{(m_1+1)}_{\alpha_{1}\dots \alpha_{m_1}j_1}(\bp_1,\dots,\bp_{m_1},\bk_1) \tilde \Gamma^{(2)}_{j_1 j_2}(-\bk_1,-\bk_2) \nonumber\\
 &\times  \tilde {\cal T}^{(m_2+1)}_{j_2\alpha_{m_1+1}\dots \alpha_{m_1+m_2}}(\bk_2,\bp_{m_1+1},\dots,\bp_{m_1+m_2})\nonumber\\
 &\equiv \int_{\{\bk_k\}} \tilde {\cal T}_1\; \tilde \Gamma^{(2)}_{j_1 j_2}(-\bk_1,-\bk_2) \; \tilde {\cal T}_2\, ,
\end{align}
where the last line defines the shorthand notations $\tilde {\cal T}_1$ and $\tilde {\cal T}_2$.
This decomposition is  represented diagrammatically on Fig. 3.
\begin{figure}
  \centering
  \begin{tikzpicture}
    \node[draw,circle,minimum size=1.3cm] (G) at (0,0) {$\tilde {\cal T}_1$};
    \node[below] at (-130:1) {$p_1$};
    \node[left,scale=1] at (-180:0.8) {$\tilde {\cal T}^{(m_1+m_2)}_{\alpha_1\dots \alpha_{m_1+m_2}}(\bp_1,\dots,\bp_{m_1+m_2}) =$};
    \draw (-130:1) -- (G);
    \draw (-100:1) -- (G);
    \node[below] at (-100:1) {$\dots$};
    \draw (-70:1) -- (G);
    \draw[-] (-40:1) -- (G);
    \draw[-] (130:1) -- (G);
    \draw[-] (100:1) -- (G);
    \node[draw,circle, pattern=north east lines] (G1) at (1.5,0) {$\tilde \Gamma^{2}$};
    \draw (G) -- (G1);
    \node[draw,circle,minimum size=1.3cm] (G2) at (3,0) {$\tilde {\cal T}_2$};
    \draw (G1) -- (G2);
     \node[below] at (2.6,-0.9) {$p_{{m_1}+1}$};
    \draw (G2) -- ++ (-100:1);
    \node[below] at (3.4,-1) {$\dots$};
    \draw (G2) -- ++ (-70:1);
    \draw (G2) -- ++ (-40:1);
    \draw (G2) -- ++ (20:1);
    \draw (G2) -- ++ (-10:1);
  \end{tikzpicture}
\caption{Decomposition of the diagram $\tilde { \cal T}$.}
\end{figure}
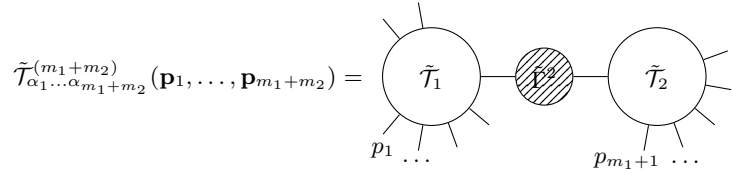
Since  $m_1+1 \leq n$ and $m_2+1 \leq n$, the property $(P)$ holds for both these diagrams.
As a consequence, taking a derivative of $\tilde {\cal T}^{(m_1+m_2)}$ with respect to $u_\mu(\bq)$ yields the following identity in the limit of large wave-numbers 
\begin{align}
 \frac{\delta}{\delta u_\mu(\bq)} &\tilde {\cal T}^{(m_1+m_2)}_{\alpha_1\dots \alpha_{m_1+m_2}}[\bp_1,\dots,\bp_{m_1+m_2};j]\Big|_{\varphi=0} \nonumber\\
 &= \int_{\{\bk_k\}} \Bigg\{-\sum_{k=1}^{m_1} \frac{p_k}{\varpi}\tilde {\cal T}^{(m_1+1)}_{\alpha_{1}\dots \alpha_{m_1} j_1}(\dots,\omega_k+\varpi,p_k,\dots,\bk_1) \tilde \Gamma^{(2)}_{j_1 j_2}(-\bk_1-\bk_2)\, \tilde { \cal T}_2\nonumber\\
 &\quad \quad -\sum_{k=m_1+1}^{m_2} \frac{p_k}{\varpi}\tilde { \cal T}_1\, \tilde \Gamma^{(2)}_{j_1 j_2}(-\bk_1-\bk_2)\tilde {\cal T}^{(m_2+1)}_{j_2\alpha_{m_1+1}\dots \alpha_{m_1+m_2}}(\bk_2,\dots,\omega_k+\varpi, p_k,\dots)\nonumber\\
 &\quad \quad - \frac{k_1^\mu}{\varpi} \tilde {\cal T}^{(m_1+1)}_{\alpha_{1}\dots \alpha_{m_1} j_1}(\bp_1,\dots,\bp_{m_1},\nu_1+\varpi,\vk_1)\tilde \Gamma^{(2)}_{j_1 j_2}(-\bk_1-\bk_2)\, \tilde { \cal T}_2\nonumber\\
 &\quad \quad -\frac{k_2^\mu}{\varpi}\tilde { \cal T}_1\, \tilde \Gamma^{(2)}_{j_1,j_2}(-\bk_1-\bk_2)\tilde {\cal T}^{(m_2+1)}_{j_2\alpha_{m_1+1}\dots \alpha_{m_1+m_2}}(\nu_2+\varpi,\vk_2,\bp_{m_1+1},\dots,\bp_{m_1+m_2})\nonumber\\
 &\quad \quad +\frac{k_1^\mu}{\varpi} \tilde {\cal T}_1\,  \tilde \Gamma^{(2)}_{j_1 j_2}(-\nu_1+\varpi,-\vk_1-\bk_2)\, \tilde {\cal T}_2 +\frac{k_2^\mu}{\varpi} \tilde {\cal T}_1\, \tilde \Gamma^{(2)}_{j_1 j_2}(-\bk_1,-\nu_2+\varpi,-\vk_2) \, \tilde {\cal T}_2 \Bigg\} \nonumber\\
& \equiv  {\cal D}_\mu(\varpi)\tilde {\cal T}^{(m_1+m_2)}_{\alpha_1\dots \alpha_{m_1+m_2}}(\bp_1,\dots,\bp_{m_1+m_2}) \, .
\end{align}
Indeed, one can easily check by making the appropriate changes of variables on the shifted frequencies that  the three last lines of the first equality cancel out by wave-vector and frequency conservation.
Furthermore, $(P)$ holds for $n=3$ because the only tree diagram with three external legs is a star diagram. Hence,  one can conclude by induction that $(P)$ holds for any trees composing the $\tilde G^{(n)}$ and by extension for their sum. Thus, the property $(P)$ is true for any connected correlation function in the limit of large wave-numbers.

\subsubsection{Action of two derivatives with respect to velocity fields}
 
 Let us then show that this property remains true when applying two derivatives.
 The key feature of the operator ${\cal D}_\mu(\varpi)$ that underlies the property $(P)$ is that
  it distributes as a functional derivative, as long as the object on which the operator  acts satisfies wave-vector
   and frequency conservation, that is symbolically
   \begin{equation}
   {\cal D}_\mu(\varpi) [F_1 F_2] = [{\cal D}_\mu(\varpi) F_1] F_2 + F_1 [{\cal D}_\mu(\varpi) F_2]  \, .
   \end{equation}
 The non-trivial aspect of this equality is that on the \lhs, the operator only acts on the external wave-vectors of the product, 
whereas on the \rhs, each operator
  acts on all the wave-vectors of each functions, including the internal ones, but these contributions cancel out 
 because of wave-vector conservation, as in the previous calculations.
  
Let us denote $\tilde {\cal T}^{(n)}_{\alpha_1\cdots\alpha_n}(\bp_1,\cdots,\bp_n)$ one of the trees composing a given correlation function $\tilde G^{(n)}$ with $n$ external wave-vectors and frequencies (of zero sum), and $\{{\cal E}_i\}$, $i=1,\dots,m$ all the elements (vertex function or propagator) composing $\tilde {\cal T}^{(n)}$:
\begin{equation}
 \tilde {\cal T}^{(n)}_{\alpha_1\cdots\alpha_n}(\bp_1,\cdots,\bp_n) = \int_{\bq_{\rm intern}}\prod_{k=1}^m {\cal E}_k[\{\bq_k\}]\,,
\end{equation}
where  $\{\bq_k\}$ denotes all the wave-vectors and frequencies attached to the element ${\cal E}_k$ (internal or external) such that when two elements are glued together by a common leg, the respective sums of the ingoing frequencies
and wave-vectors carried by the corresponding legs are zero and the integration is only on internal wave-vectors and frequencies.

The property $(P)$ amounts to have 
\begin{equation}
 {\cal D}_\mu(\varpi) \int_{\bq_{\rm intern}} \prod_{k=1}^m {\cal E}_k[\{\bq_k\}] = 
 \int_{\bq_{\rm intern}} \sum_{k=1}^m \prod_{i =1 \atop i\neq k}^m {\cal E}_i[\{\bq_i\}]  {\cal D}_\mu(\varpi){\cal E}_k[\{\bq_k\}]\, .
\label{eq:B1}
  \end{equation}
We now show that this property can be generalized to two derivatives. The two functional derivatives acting on $\tilde T^{(n)}$
 can be expressed as
\begin{align}
\frac{\delta^2 }{\delta u_{\mu_1}(\bq_1) \delta u_{\mu_2}(\bq_2)} \int_{\bq_{\rm intern}} \prod_{k=1}^m {\cal E}_k[\{\bq_k\}]\Big|_{\varphi=0} &= \int_{\bq_{\rm intern}}\sum_{i=1}^m\sum_{j=1 \atop j \neq i}^m
\prod_{k=1 \atop {k\neq i \atop k\neq j}}^m {\cal E}_k[\{\bq_k\}]\left(\frac{\delta}{\delta u_{\mu_1}(\bq_1)}{\cal E}_i[\{\bk_i\}]\right)
\left (\frac{\delta}{\delta  u_{\mu_2}(\bq_2)}{\cal E}_j[\{\bk_j\}]\right)\nonumber\\
&+\int_{\bq_{\rm intern}}\sum_{i=1}^m
\prod_{k=1 \atop {k\neq i \atop }}^m {\cal E}_k[\{\bq_k\}]\left(\frac{\delta^2}{\delta u_{\mu_1}(\bq_1) \delta  u_{\mu_2}(\bq_2)}{\cal E}_i[\{\bk_i\}]\right)\, .
\label{eq:B2}
\end{align}
If each of the derivatives act on a different element, then, the property $(P)$ applies to both elements locally,
  and the functional derivative can be replaced by a ${\cal D}_{\mu}(\varpi)$ operator.
If the two derivatives act on the same element, this is also the case.
Indeed, if the element is a vertex, one has in the limit of large wave-numbers
 \begin{align}
 \frac{\delta^2}{\delta u_{\mu_1}(\bq_1)\delta u_{\mu_2}(\bq_2)}\tilde \Gamma_{\alpha_1\dots\alpha_{k+\ell}}^{(k,\ell)}[\bp_1,\dots,\bp_{k+\ell};\varphi] \Big|_{\varphi=0}
 &\underset{|\vp_i|\gg \kappa}{=}  \tilde \Gamma_{\mu_1{\mu_2}\alpha_1\dots\alpha_{k+\ell}}^{(k+2,\ell)}(\varpi_1,\vq_1=0,\varpi_2,\vq_2=0,\bp_1,\dots,\bp_{k+\ell})\nonumber\\
& \quad =  \sum_{i=1}^{k+\ell}\frac{p_i^{\mu_1}}{\varpi_1}\sum_{j=1}^{k+\ell}\frac{p_j^{{\mu_2}}}{\varpi_2}
\;\;\tilde \Gamma^{(k,\ell)}_{\alpha_1\dots\alpha_{k+\ell}}(\cdots,\omega_i+\varpi_1,\vp_i,\cdots,\omega_j+\varpi_2,\vp_j\cdots)\nonumber\\
 &\quad \equiv{\cal D}_{\mu_1}(\varpi_1) {\cal D}_{\mu_2}(\varpi_2)\tilde \Gamma_{\alpha_1\dots\alpha_{k+\ell}}^{(k,\ell)}(\bp_1,\dots,\bp_{k+\ell})\, , \nonumber\\
 \end{align} 
 applying twice the Ward identity \eq{eq:wardGalN}.
 If the element is a propagator, one obtains
\begin{align}
 &\frac{\delta^2}{\delta  u_{\mu_1}(\bq_1)\delta u_{\mu_2}(\bq_2)} 
 \tilde G^{(2)}_{\alpha_1 \alpha_2}[\bp_1,\bp_2;j]\Big|_{\varphi=0}  \nonumber\\
 & \quad = - \int_{ \bk_1,\bk_2} \tilde G^{(2)}_{\alpha_1 i}(\bp_1,-\bk_1) \tilde \Gamma^{(4)}_{i j {\mu_1}{\mu_2}} (\bk_1,\bk_2,\bq_1,\bq_2) \tilde G^{(2)}_{j\alpha_2}(-\bk_2,\bp_2)\nonumber\\
 & \quad\quad + \int_{ \bk_i} \tilde G^{(2)}_{\alpha_1 i}(\bp_1,-\bk_1) \tilde \Gamma^{(3)}_{i j {\mu_1}} (\bk_1,\bk_2,\bq_1)
 \tilde G^{(2)}_{j k}(-\bk_2,-\bk_3) \tilde \Gamma^{(3)}_{k\ell  {\mu_2}} (\bk_3,\bk_4,\bq_2)
   \tilde G^{(2)}_{\ell\alpha_2}(-\bk_4,\bp_2)+ (\mu_1,\bq_1) \longleftrightarrow  (\mu_2,\bq_2) \nonumber\\
  &\quad\hspace{-0.3cm}\underset{|\vp_i|\gg \kappa}{=} - 
 \int_{ \bk_1,\bk_2} \tilde G^{(2)}_{\alpha_1 i}(\bp_1,-\bk_1)\Big[ \frac{k_1^{\mu_1} k_1^{\mu_2}}{\varpi_1\varpi_2}\tilde \Gamma^{(2)}_{i j} (\nu_1+\varpi_1+\varpi_2,\vk_1,\bk_2) +\frac{k_1^{\mu_1} k_2^{\mu_2}}{\varpi_1\varpi_2} \tilde \Gamma^{(2)}_{i j} (\nu_1+\varpi_1,\vk_1,\nu_2+\varpi_2,\vk_2)\nonumber\\
 &\quad\quad +\frac{k_2^{\mu_1} k_1^{\mu_2}}{\varpi_1\varpi_2} \tilde \Gamma^{(2)}_{i j} (\nu_1+\varpi_2,\vk_1,\nu_2+\varpi_1,\vk_2) +\frac{k_2^{\mu_1} k_2^{\mu_2}}{\varpi_1\varpi_2}\tilde \Gamma^{(2)}_{i j} (\bk_1,\nu_2+\varpi_1+\varpi_2,\vk_2)
 \Big]\tilde G^{(2)}_{j\alpha_2}(-\bk_2,\bp_2)  \nonumber\\
 &\quad\quad +\int_{ \bk_i} \tilde G^{(2)}_{\alpha_1 i}(\bp_1,-\bk_1) \Big[\frac{k_1^{\mu_1}}{\varpi_1}\tilde \Gamma^{(2)}_{i j} (\nu_1+\varpi_1,\vk_1,\bk_2)+
 \frac{k_2^{\mu_1}}{\varpi_1}\tilde \Gamma^{(2)}_{i j} (\bk_1,\nu_2+\varpi_1,\vk_2)\Big]\tilde G^{(2)}_{j k}(-\bk_2,-\bk_3) \nonumber\\
 &\quad\quad\times  \Big[ \frac{k_3^{\mu_2}}{\varpi_2}\tilde \Gamma^{(2)}_{k \ell} (\nu_3+\varpi_2,\vk_3,\bk_4)+
 \frac{k_4^{\mu_2}}{\varpi_2}\tilde \Gamma^{(2)}_{k \ell} (\bk_3,\nu_4+\varpi_2,\vk_4)\Big] \tilde G^{(2)}_{\ell\alpha_2}(-\bk_4,\bp_2) + (\mu_1,\bq_1) \longleftrightarrow  (\mu_2,\bq_2)\nonumber\\
 &\quad= \frac{p_1^{\mu_1} p_1^{\mu_2}}{\varpi_1\varpi_2}\tilde G^{(2)}_{\alpha_1\alpha_2}(\omega_1+\varpi_1+\varpi_2,\vp_1,\bp_2) +\frac{p_1^{\mu_1} p_2^{\mu_2}}{\varpi_1\varpi_2}\tilde G^{(2)}_{\alpha_1\alpha_2}(\omega_1+\varpi_1,\vp_1,\omega_2+\varpi_2,\vp_2)\nonumber\\
 &\quad\quad +\frac{p_2^{\mu_1} p_2^{\mu_2}}{\varpi_1\varpi_2}\tilde G^{(2)}_{\alpha_1\alpha_2}(\bp_1,\omega_2+\varpi_1+\varpi_2,\vp_2) +\frac{p_2^{\mu_1} p_1^{\mu_2}}{\varpi_1\varpi_2}\tilde G^{(2)}_{\alpha_1\alpha_2}(\omega_1+\varpi_2,\vp_1,\omega_2+\varpi_1,\vp_2)\nonumber\\
 &\quad= {\cal D}_{\mu_1}(\varpi_1) {\cal D}_{\mu_2}(\varpi_2) \tilde G^{(2)}_{\alpha_1 \alpha_2}(\bp_1,\bp_2)\, .
\end{align} 
Thus, in the limit of large wave-numbers, one deduces that \Eq{eq:B1} can be expressed as 
\begin{align}
\frac{\delta^2 }{\delta u_{\mu_1}(\bq_1) \delta u_{\mu_2}(\bq_2)} \int_{\bq_{\rm intern}}\prod_{k=1}^m {\cal E}_k[\{\bq_k\}]\Big|_{\varphi=0} 
&= \int_{\bq_{\rm intern}}\sum_{i=1}^m\sum_{j=1 \atop j \neq i}^m
\prod_{k=1 \atop {k\neq i \atop k\neq j}}^m {\cal E}_k[\{\bq_k\}]\Big({\cal D}_{\mu_1}(\varpi_1){\cal E}_i[\{\bq_i\}]\Big)
\Big({\cal D}_{\mu_2}(\varpi_2){\cal E}_j[\{\bq_j\}]\Big)\nonumber\\
&\quad+\int_{\bq_{\rm intern}}\sum_{i=1}^m
\prod_{k=1 \atop {k\neq i \atop }}^m {\cal E}_k[\{\bq_k\}]\left({\cal D}_{\mu_1}(\varpi_1){\cal D}_{\mu_2}(\varpi_2){\cal E}_i[\{\bk_i\}]\right)\nonumber\\
 &= \int_{\bq_{\rm intern}}{\cal D}_{\mu_1}(\varpi_1) \sum_{j=1}^m \prod_{k=1 \atop k\neq j}^m {\cal E}_k[\{\bq_k\}]\Big({\cal D}_{\mu_2}(\varpi_2){\cal E}_j[\{\bq_j\}]\Big) \nonumber\\
 &= {\cal D}_{\mu_1}(\varpi_1){\cal D}_{\mu_2}(\varpi_2) \int_{\bq_{\rm intern}} \prod_{k=1}^m {\cal E}_k[\{\bq_k\}] \, , 
\end{align}
where in the last two lines the property \eq{eq:B2} has been used since wave-vector and frequency conservation is satisfied for each element ${\cal E}_i$ and also for the modified elements ${\cal D}_{\mu}(\varpi){\cal E}_j$ with a frequency shift.
One concludes that taking two functional derivatives  of $\tilde G^{(n)}$ is equivalent at large wave-numbers to applying twice the finite difference operator ${\cal D}$, such that the exact leading term of the flow equation for any correlation function can be expressed in a closed form, and reads
\begin{equation}
\p_\kappa \bar G^{(n)}_{\alpha_1\dots\alpha_{n}}({\bp_1}, \dots, {\bp_{n-1}}) = 
 \frac{1}{2}\,  \int_{\bq}\! \tilde \p_\kappa \bar G^{(2,0)}_{\mu \nu}(\bq) 
  {\cal D}_\mu(\varpi){\cal D}_\nu(-\varpi) \bar G^{(n)}_{\alpha_1\dots\alpha_{n}}({\bp_1}, \dots, {\bp_{n-1}})\, .
\label{eq:flowGNmunu}
\end{equation}

Finally, let us explicitly use the transversality of the $2-$point function. We note $\bar G^{(2,0)}_{\mu\nu}({\bq}) =P_{\mu\nu}^\perp(\vq) \bar C_\kappa(q,\varpi)$
 where the transverse projector is defined by
 $P_{\mu\nu}^\perp(\vq)=\delta_{\mu\nu}-{q_\mu q_\nu}/{q^2}\, .$
The angular integration in $\vq$ can be performed, using $\int_{\vq} q_\mu q_\nu f(q) = \frac{\delta_{\mu\nu}}{d} \int_{\vq} q^2 f(q)$.
 One obtains as a final result  that the leading contribution to the flow equation at large wave-numbers is
\begin{equation}
\p_\kappa \bar G^{(n)}_{\alpha_1\dots\alpha_{n}}({\bp_1}, \dots, {\bp_{n-1}}) = 
 \frac{d-1}{2d}\,  \int_{\bq}\! \tilde \p_\kappa \bar C(\varpi,\vq) 
  {\cal D}_\mu(\varpi){\cal D}_\mu(-\varpi) \bar G^{(n)}_{\alpha_1\dots\alpha_{n}}({\bp_1}, \dots, {\bp_{n-1}})\, .
\label{eq:flowGNapdx}
\end{equation}
 Because of the properties of the BMW approximation,  the sub-leading contributions in this equation can be bounded in the following way \cite{Benitez12}
\begin{equation}
 \p_\kappa \bar G^{(n)} = \p_\kappa \bar G^{(n)}\Big|_{\rm leading} + R^{(n)} 
 \quad\quad\hbox{with}\quad\quad R^{(n)}= {\cal O}\left(\frac{\kappa}{p_{\rm min}}\right)\p_\kappa \bar G^{(n)}\Big|_{\rm
  leading}\, ,\label{eq:bound}
\end{equation}
where $p_{\rm min}$ is the minimum of the moduli of the 
 $\vp_i$ and of their partial sums, and $\p_\kappa \bar G^{(n)}\Big|_{\rm leading}$ is the r.h.s of \Eq{eq:flowGNapdx}.

\section{Solution of the fixed-point equations}
\label{AP:solution}
The aim of this Appendix is to derive the solution of flow equation \eq{eq:flowGNapdx} for $n$-point correlation functions at the fixed point.
To study the fixed point, one introduces dimensionless quantities, denoted by a hat symbol. 
Wave-numbers are measured in units of $\kappa$, \eg $p = \kappa \hat p$.
 The dimensionless effective viscosity  $\nu_{\kappa}$ and effective forcing $D_{\kappa}$ are defined, as in \aref{AP:RG}, as
 \begin{equation} 
 N_\kappa(\vp) \equiv D_\kappa \hat n(\hp) \quad\quad\quad \hbox{and}\quad\quad\quad  R_\kappa(\vp) \equiv \nu_\kappa \hat r(\hp) \, . 
 \end{equation}
  One deduces that times are measured in units of $(\kappa^2 \nu_\kappa)^{-1}$.
 The coefficient $D_\kappa$ is related to the mean energy injection rate $\varepsilon$ as
  \begin{equation}
 \varepsilon = \langle f_\alpha(t,\vx)v_\alpha(t,\vx)  \rangle =
  D_\kappa \kappa^d \, (d-1)\int_{\hat \varpi,\hat\vq} \hat n(\hat q) \bar{G}(\hat\omega,\hat q) \equiv D_\kappa  \kappa^d \gamma^{-1} \, ,
 \end{equation}
 using Janssen de Dominicis formalism to express the average value of a quantity linear in the forcing 
   in term of the response field ($\bar{G}$ being the transverse part of the response function $\bar{G}^{(1,1)}$), and where 
 $\gamma$ is a non-universal number, depending on the forcing profile through $\hat n$ \cite{Canet16}.
 This allows one to express $D_\kappa$ as
 \begin{equation}
 D_\kappa = \varepsilon \gamma \kappa^{-d}\, .
 \end{equation}
 The running coefficient $\nu_\kappa$ is expected to behave at the fixed point as a power-law
  $\nu_\kappa\sim \kappa^{-4/3}$ in $d=3$, where the exponent is fixed by Galilean invariance \cite{Canet16}.
   One deduces that
   \begin{equation}
    \nu_\kappa = \nu_{\eta^{-1}} ({\kappa}\eta)^{-4/3} \simeq \nu_{\Lambda} ({\kappa}\eta)^{-4/3} = \epsilon^{1/3} \kappa^{-4/3}\, ,
   \end{equation}
  assuming that the fixed-point is already attained at scale $\eta^{-1}$ and  neglecting the evolution of $\nu_\kappa$ between 
   the microscopic scale $\Lambda$ and  $\eta^{-1}$.
 Hence we define dimensionless times  as $\hat t = \epsilon^{1/3} \kappa^{2/3} t$.
  The definitions of the dimensionless velocity and response velocity fields then follow as
 \begin{equation}
  v_\alpha = (\kappa^{d-2} D_\kappa \nu_\kappa^{-1} )^{1/2} \hat v_\alpha = \kappa^{-1/3}\epsilon^{1/3}\gamma^{1/2} \, \hat v_\alpha \quad\quad\quad \hbox{and}\quad\quad\quad
  \bar v_\alpha = (\kappa^{d+2} \nu_\kappa D_\kappa^{-1} )^{1/2}\, \hat{\bar {v}}_\alpha =\kappa^{10/3}\epsilon^{-1/3}\gamma^{-1/2}\, \hat{\bar {v}}_\alpha\, . \label{eq:dim}
 \end{equation}
Let us finally define the dimensionless integral $\hat J_s$  through
\begin{equation}
J_\kappa(\varpi) = - \int_\vq \tilde{\p}_s{\bar{C}}(\varpi,\vq)  = \gamma \epsilon^{1/3}\kappa^{-4/3}\hat J_s(\hat \varpi)\, ,
\end{equation}
 where  $s \equiv \ln{(\kappa/\Lambda)}$ is the "\RG time" introduced in \aref{AP:RG}.

\subsection{Solution of the fixed-point equation for the $2-$point function for large time delays}

In this section, we derive the fixed-point solution of \Eq{eq:flowCt} at large time delays.
For this, let us write in \Eq{eq:flowCt} $J_\kappa$ as $J_\kappa(\varpi ) = (J_\kappa(\varpi )-J_\kappa(0))+J_\kappa(0)$.
Since $J_\kappa$ is a regular even function of $\varpi$,
  the term $F(\varpi)=(J_\kappa(\varpi )-J_\kappa(0))/\varpi ^2$ is an analytic function of $\varpi$.
 It follows that its Fourier transform $\int \cos(\varpi t) F(\varpi)$  decays exponentially in $t$,
 and its integral $\int F(\varpi)$ is a constant independent of $t$.
 At large $t$, the integral in \eq{eq:flowCt} is thus dominated by the remaining term which turns out to be of order $t$: 
\begin{equation}
J_\kappa(0) \int_{-\infty}^\infty\frac{d\varpi}{2\pi}\, \frac{\cos(\varpi t)-1}{\varpi^2} = -\frac{J_\kappa(0)}{2} |t|.
\end{equation}
The flow equation \Eq{eq:flowCt} hence reduces in the limit $t\gg \kappa^{2/3}$ to 
\begin{equation}
\kappa \partial_\kappa  C(t, \vp)= \frac{J_\kappa(0)}{3} \,|t|\, p^2 \,C(t,\vp).
\end{equation}
Using \Eq{eq:dim},  the dimensionless 2-point correlation function $\hat C_s(\hat t,\hat p)$ can be defined as
\begin{equation}
 C_\kappa(t,\vp) =  \frac{\gamma \epsilon^{2/3}}{p^{11/3}}  \hat C_s\left(\hat y = 
 \epsilon^{1/3}\,t p^{2/3}, \hat p ={p}/{\kappa}\right)\, ,
\end{equation}
 and one obtains the dimensionless flow equation
\begin{equation}
 \p_s \hat C_s(\hy,\hp)  - \hp \p_{\hp} \hat C_s(\hy,\hp) = 
 \frac{\hat J_s(0)}{3}\,\gamma \hp^{4/3} \, |\hat y| \hat C_s(\hy,\hp) \, .
\label{eq:Xfix}
\end{equation}
The fixed point corresponds  to setting $\p_s \hat C_s=0$ as explained in Appendix (\ref{AP:RG}), and  
$\hat J_s(0) \to \hat J_*$, $\hat C_s(\hy,\hp) \to \hat C_*(\hy,\hp)$ tend to fixed point quantities.
The solution of the fixed-point equation is thus given by
\begin{equation}
 \log \hat C_*(\hy,\hp) = -\frac{\gamma \hat J_*}{6} \hp^2 |\hat t| + \hat F(\hy) + {\cal O}(\hat p)\, ,
\end{equation}
where $\hat F$ is a regular function of $\hy$  and the error term is calculated from \Eq{eq:bound} (see below).
The leading term explicitly breaks scale invariance. Because of this violation, the fixed point solution explicitly depends  on a scale $\kappa$, when expressed in terms of the dimensionful variables. 
 The physical correlation functions are obtained in the limit $\kappa\to 0$. Since the flow essentially stops when 
 passing the inverse integral scale (the fixed point is attained and by definition the dimensionless quantities stop evolving), the relevant scale is of the order of $L^{-1}$.
  The physical (dimensionful) solution,  indexed by the subscript '$L$' for 'long' time, is thus given by
  \begin{equation}
  \log \Big[\frac{C_L(t,\vp)}{\varepsilon^{2/3}L^{11/3}} \Big] = - \alpha_{L} \varepsilon^{1/3} L^{4/3} |t|\, p^2
   - \frac{11}{3} \log (pL) + F_L(\varepsilon^{1/3}p^{2/3}t) + {\cal O}(pL)\,\, , 
  \end{equation}
with $\alpha_{L} ={\hat J_*}\gamma/6$ a non-universal constant. Note that  we have chosen  to keep sub-leading
terms in the solution, although they are of the same order as the error, because they embody the Kolmogorov scaling solution:
 \begin{equation}
 C_K(t,\vp) \propto \frac{\varepsilon^{2/3}}{p^{11/3}}  \,H_K(\varepsilon^{1/3} p^{2/3}t)
 \end{equation}
where $H_K=\exp(F_L)$ is a scaling function. However, as explained in the main text, this part is not calculated exactly,
 and could receive corrections, which are ${\cal O}(pL)$, from the sub-leading terms in the flow equation, neglected here.

\subsection{Solution of the fixed point  equation for a $n-$point function}

In this section, we determine the fixed point solution of the leading contribution to the flow equation \eq{eq:lowGt}
 for the hybrid time-wave-vector generalized $n$-point correlation function $G^{(n)}$, which reads
\begin{equation}
 \p_\kappa  G^{(n)}_{\alpha_1\dots\alpha_{n}} ({t_1,\vp_{1}},\cdots,{t_{n-1},\vp_{n-1}}) = \frac{1}{3}\,   G^{(n)}_{\alpha_1\dots\alpha_{n}}  ({t_1,\vp_{1}},\cdots,{t_{n-1},\vp_{n-1}})
 \sum_{k,\ell} \vp_k \cdot \vp_\ell \int_{\varpi}\! J_\kappa(\varpi)\,
\frac{e^{i\varpi(t_k - t_\ell )} - e^{i\varpi t_k} - e^{-i\varpi t_\ell} +1}{\varpi^2} \, .
\label{eq:dup}
\end{equation}

\subsubsection{Small time delays}

If one defines $t_i \equiv \epsilon\, \tilde t_i$ and let $\epsilon$ tend to zero, the integrals in the \rhs of \Eq{eq:dup} are equivalent
 to $ \epsilon^2 I_{\kappa} \tilde t_k \tilde t_\ell $, where $I_{\kappa} \equiv \int_{\varpi}\! J_\kappa(\varpi)$:
\begin{equation}
 \lim_{\epsilon \to 0} \frac{1}{\epsilon^2}\! \int_{\varpi}\! J_\kappa(\varpi)
 \frac{e^{i\varpi(t_k - t_\ell )} - e^{i\varpi t_k} - e^{-i\varpi t_\ell } +1}{\varpi^2} =I_{\kappa} \tilde t_k \tilde t_\ell \, .
 \label{eq:flowAp}
\end{equation}
 Furthermore, it was shown in \citep{Canet16} that, because of the presence of the regulator, $\hat J_\kappa(\varpi)$
  is dominated by frequencies of order $\kappa^{2/3}$, and that $I_{\kappa}$ is finite.
In the limit where all the time delays $t_i$ are small, the flow equation \eq{eq:dup} simplifies to 
\begin{align}
 \p_\kappa  G^{(n)}_{\alpha_1\dots\alpha_{n}}({t_1,\vp_1}, \cdots ,{t_{n-1},\vp_{n-1}})
 &=  \frac{I_{\kappa}}{3} |\vp_k t_k|^2  G^{(n)}_{\alpha_1\dots\alpha_{n}}({t_1,\vp_1}, \cdots, {t_{n-1},\vp_{n-1}})\, ,
\end{align}
(using Einstein convention for repeated indices). In order to find a solution, we define a $(n-1)\times(n-1)$ rotation matrix ${\cal R}$, such that ${\cal R}_{i1} = \frac{t_i}{\sqrt{t_\ell t_\ell}}$, and introduce new variables $\vec{\rho}_k$ such that
$\vp_i = {\cal R}_{ij}\vec{\rho}_j$. In particular $\vec{\rho}_1 = \frac{t_k \vp_k}{\sqrt{t_\ell t_\ell}}$ and the flow equation becomes
\begin{align}
 \p_\kappa  G^{(n)}_{\alpha_1\dots\alpha_{n}}({t_1,\vec{\rho}_1}, \cdots, {t_{n-1},\vec{\rho}_{n-1}})
 &= \frac{I_{\kappa}}{3}  t_k t_k|\vec{\rho}_1|^2 \bar G^{(n)}_{\alpha_1\dots\alpha_{n}}({t_1,\vec{\rho}_1} ,\cdots ,{t_{n-1},\vec{\rho}_{n-1}})\, .
\end{align}

To study the fixed point of this flow equation, we introduce the  dimensionless variables
$\hat{\vec{\rho}}_i \equiv \vec{\rho}_i/\kappa$, and 
 $\hat I_{s} \equiv \gamma^{-1} \varepsilon^{-2/3}\kappa^{2/3} I_{\kappa}$. According to  \Eq{eq:dim}, the dimensionless
  $n-$point function can be defined as
 \begin{align}
 \hat{{G}}^{(n)}_{\alpha_1\dots\alpha_{n}}({\hat t_1,\hat{\vec{\rho}}_1}, \cdots, {\hat t_{n-1},\hat{\vec{\rho}}_{n-1}}) &\equiv
  \Big(\frac{\kappa^{11/3}}{\gamma \varepsilon^{2/3}}\Big)^{\frac{m-\bar{m}}{2}} \kappa^{\frac{3}{2}(m + \bar{m} -2)}\,
    G^{(n)}_{\alpha_1\dots\alpha_{n}}({t_1,\vec{\rho}_1}, \cdots, {t_{n-1},\vec{\rho}_{n-1}}) \, ,
 \end{align}
  where $m$ (resp. $\bar{m}$) is the number of velocity (resp. response velocity) fields 
  in the generalized correlation function $\hat{{G}}^{(n)}$, 
   with $m+ \bar m =n$. We note $d_G = 3(m-1) + (m -\bar{m})/3$ the scaling dimension of $G^{(n)}$,
 and we define $\alpha_{s} = \gamma \hat I_{s}/2$ (which fixed point value $\alpha_*\equiv\alpha_S$ is the coefficient appearing in \eq{eq:solC}).
The flow equation for $\hat{{G}}^{(n)}$ then reads
\begin{equation}
 \Big\{\p_s -d_G - \hat{\vec{\rho}}_i \cdot \p_{\hat{\vec{\rho}}_i} + \frac{2}{3} \hat t_i \p_{\hat t_i}
  -\frac{2}{3}\alpha_{s}\, \hat t_k \hat t_k|\hat{\vec{\rho}}_1|^2\Big\}
 \hat{{G}}^{(n)}_{\alpha_1\dots\alpha_{n}}({\hat t_1,\hat{\vec{\rho}}_1}, \cdots, {\hat t_{n-1},\hat{\vec{\rho}}_{n-1}}) =0\, .
\end{equation}
In the following, hat symbols are omitted to alleviate notation. Let us  remark that $\vec{\rho}_i \cdot \p_{\vec{\rho}_i}$ only acts on the moduli of the vectors $\vec{\rho}_i$, so that 
 if one introduces the polar decomposition $\vec{\rho}_i = \rho_i \vec{n}_i$, one has
$\vec{\rho}_i \cdot \p_{\vec{\rho}_i} = \rho_i \p_{\rho_i}$. Introducing the scaling variables $y_i = {\rho_1}^{2/3} t_i$, the flow equation further simplifies  to
\begin{equation}
 \Big\{\p_s -d_G - \rho_i \p_{\rho_i} -\frac{2}{3}\alpha_{s}\, y_k y_k \rho_1^{2/3}\Big\}
 {G}^{(n)}_{\alpha_1\dots\alpha_{n}}({y_1,\rho_1,\vec{n}_1}, \cdots, {y_{n-1},\rho_{n-1},\vec{n}_{n-1}}) =0\, .
\end{equation}
Let us consider \RG scales $s$ such that the fixed point is reached: $\alpha_s$ has attained its fixed point value $\alpha_*$, and  the explicit dependence in $s$ (through $\p_s$) is zero.
 Denoting $u_1 \equiv \ln{\rho_1},\; u_{i > 1} \equiv \ln{\rho_1} - \ln{\rho_i}$, the fixed point equation becomes an ordinary differential equation 
\begin{equation}
 \Big\{-d_G - \p_{u_1} -\frac{2}{3}\alpha_{*}\,  y_k y_k\, e^{\frac{2}{3} u_1}\Big\} 
 {G}^{(n)}_{\alpha_1\dots\alpha_{n}}({y_1,u_1,\vec{n}_1}, \cdots ,{y_{n-1},u_{n-1},\vec{n}_{n-1}}) =0\, .
\label{eq:diffGn}
\end{equation}
 For this equation, the bound \eq{eq:bound} on the sub-leading contributions is given by
\begin{equation}
 R^{(n)} = {\cal O}\left(\frac{\kappa}{p_{\rm min}}\right) t_k t_k \rho_1^2 {G}^{(n)} = {\cal O}\left(\kappa p_{\rm max}  \right) {G}^{(n)}
\label{eq:boundGn}
\end{equation}
The differential equation \Eq{eq:diffGn} can be integrated, and yields
\begin{equation}
 \log {G}^{(n)}_{\alpha_1\dots\alpha_{n}}({y_1,u_1,\vec{n}_1}, \cdots, {y_{n-1},u_{n-1},\vec{n}_{n-1}}) 
 = -\alpha_{*} \,  y_k y_k\, e^{\frac{2}{3}u_1} - d_G u_1 + F^{(n)}_{\alpha_1\dots\alpha_{n}}({y_1,\vec{n}_1}, \cdots, {y_{n-1},u_{n-1},\vec{n}_{n-1}}) \, .
 \end{equation}
 In terms of the  original dimensionful variables, one obtains
\begin{align}
 \log \Big[\varepsilon^{\frac{\bar{m}-m}{3}}L^{-d_G}{G}^{(n)}_{\alpha_1\dots\alpha_{n}}({t_1,\vp_1}, \cdots, {t_{n-1},\vp_{n-1}})\Big] &= 
 -\alpha_S\varepsilon^{2/3}L^{2/3} \,  t_k t_k\, \rho_1^2\nonumber\\
 - d_G \log (\rho_1 L) &+ {{F_S}^{(n)}}_{\alpha_1\dots\alpha_{n}}\left(\rho_1^{2/3}\varepsilon^{1/3} t_1,\frac{\vec{\rho}_{1}}{\rho_1}, \cdots,
 \rho_1^{2/3}\varepsilon^{1/3}t_{n-1}, \frac{\vec{\rho}_{n-1}}{\rho_1}\right) + {\cal O}(p_{\rm max} L)\, ,
\label{eq:solGnpetit}
\end{align}
where $\vec{\rho}_1 = \frac{t_k \vp_k}{\sqrt{t_\ell t_\ell}}$  and the dimensionless constant
 $\gamma$ has been absorbed in the function $F_S^{(n)}$. The error on the solution can be simply deduced
  using \Eq{eq:boundGn}: it is bounded by a term of order $p_{\rm max} L$, where $p_{\rm max}$ is the maximum 
  of the amplitudes of the $\vp_i$ and of the partial sums.  Note that again the Kolmogorov solution, which stems from standard
 scale invariance, is included 
  explicitly although it is of the same order as the neglected error terms. This part is not calculated exactly
   since it could receive corrections from the neglected sub-leading terms in the flow equation.
The leading term in $\bar G^{(n)}$  is a
 Gaussian in the variable $|\vp_k t_k|$, which explicitly breaks scale invariance. This breaking is related
  to  the sweeping effect, and expression \eq{eq:solGnpetit} provides its exact expression, as a generalization
   of the Gaussian in $tp$ for the two-point function.
The constant $\alpha_S$ is positive, see \citep{Canet16}.
 The vectors $\vec{\rho}_i$ are  not given explicitly, except for $\vec{\rho}_1$, but they can be constructed for any generalized correlation function.
 
For instance, in the  case of $G^{(3)}$, one can use  $\mathcal{R} \propto \big( \begin{smallmatrix} t_1&-t_2\\ t_2&t_1 \end{smallmatrix} \big)$ as the rotation matrix
 from the $\vec{\rho}_i$ to the $\vp_i$, so that
\begin{align}
 &\log \big[\varepsilon^{-1} L^{-7} {G}^{(3)}_{\alpha\beta\gamma}({t_1,\vp_1}, {t_2,\vp_{2}})\big] = 
 -\alpha_{S}\varepsilon^{2/3} L^{2/3}\,  \abs{\vp_1t_1 + \vp_2t_2}^2 -7\log \Bigg( L\frac{\abs{\vp_1t_1 + \vp_2t_2}}{\sqrt{t_1^2 + t_2^2}}\Bigg) \nonumber\\
 &\quad\quad + {F_S^{(3)}}_{\alpha\beta\gamma}\Bigg(
 {\frac{\abs{\vp_1t_1 + \vp_2t_2}}{(t_1^2 + t_2^2)^{1/3}}}^{2/3} \varepsilon^{1/3}t_1, {\frac{\vp_1t_1 + \vp_2t_2}{\abs{\vp_1t_1 + \vp_2t_2}}},
 {\frac{\abs{\vp_1t_1 + \vp_2t_2}}{(t_1^2 + t_2^2)^{1/3}}}^{2/3} \varepsilon^{1/3}t_2, 
 {\frac{\vp_2t_1 - \vp_1t_2}{\abs{\vp_1t_1 + \vp_2t_2}}}\Bigg) + {\cal O}(p_{\rm max}L)\, .
\label{eq:solGn}
\end{align}
 A particular and interesting case corresponds to $t_1 = t_2 = t$. In this case, the expression 
 (\ref{eq:solGn}) simplifies to
\begin{align}
 &\log \big[\varepsilon^{-1} L^{-7}{G}^{(3)}_{\alpha\beta\gamma}({t,\vp_1}, {t,\vp_{2}})\big] = -\alpha_{S}\varepsilon^{2/3}
  L^{2/3}\,  t^2 \abs{\vp_1 + \vp_2}^2 +  {\cal O}(p_{\rm max}L)\, ,
\label{eq:solGnTeq}
\end{align}
omitting the Kolmogorov terms. This simple prediction could be tested in numerical simulations of \NS equation.

\subsubsection{Large time delays}

In this section, we derive the form of the fixed point solution for the flow equation \eq{eq:lowGt}
in the limit of large time delays.
 In this limit, one can reiterate the calculation for the two-point function,
 writing $J_\kappa(\varpi) = (J_\kappa(\varpi)-J_\kappa(0))+ J_\kappa(0)$. The integrals of the type $\int_\varpi e^{i\varpi t} F(\varpi)$
 decay exponentially in $t$ since $F(\varpi) =(J_\kappa(\varpi)-J_\kappa(0))/\varpi^2$ is an analytic function of $\varpi$, while $\int_\varpi F(\varpi)$ is a constant. Thus, the integral in  \eq{eq:lowGt} in dominated at large times  by 
\begin{align}
  J_\kappa(0) \int_\varpi \frac{e^{i\varpi(t_k - t_\ell )} - e^{i\varpi t_k} - e^{-i\varpi t_\ell} +1}{\varpi^2}
&= 4 J_\kappa(0) \int_{-\infty}^{\infty} \frac{d\varpi}{2\pi} \sin\left(\frac{\varpi t_k}{2}\right)\sin\left(\frac{\varpi t_\ell}{2}\right)
 \frac{e^{i\frac{\varpi}{2}(t_k-t_\ell)}}{\varpi^2}\nonumber\\
& = \frac{J_\kappa(0)}{2} \Big( |t_k| +|t_\ell| -|t_k-t_\ell| \Big)\, .
\end{align}
The flow equation \Eq{eq:lowGt} thus reads in the limit of large time delays $t_i\gg \kappa^{2/3}$
\begin{equation}
 \p_\kappa G^{(n)}_{\alpha_1\dots\alpha_{n}}(\{t_i,\vp_i\})=\frac{J_{\kappa}(0)}{6}\,
  \sum_{k,\ell} \vp_k\cdot\vp_\ell \Big( |t_k| +|t_\ell| -|t_k-t_\ell| \Big)\bar G^{(n)}_{\alpha_1\dots\alpha_{n}}(\{t_i\,,\vp_i\}) \, .
\end{equation}

  To give an important concrete example, let us  focus on the special case where
  all time differences are equal $t_i\equiv t$ for $i=1,\cdots, n-1$.
 In this case, the solution can be simply derived.
Introducing as previously a $(n-1)\times(n-1)$ matrix ${\cal R'}$, such that ${\cal R'}_{i1} = 1$, one defines  
the variables $\vec{\varrho}_k$ by
$\vp_i = {\cal R'}_{ij}\vec{\varrho}_j$ with $\vec{\varrho}_1 =  \sum_k \vec p_k$. The flow equation becomes
\begin{align}
 \p_\kappa  G^{(n)}_{\alpha_1\dots\alpha_{n}}({t,\vec{\varrho}_1}, \cdots,\vec{\varrho}_{n-1})
 = \frac{J_{\kappa}(0)}{3}\,|\vec{\varrho}_1|^2 \,|t| \, \bar G^{(n)}_{\alpha_1\dots\alpha_{n}}({t,\vec{\varrho}_1} ,\cdots ,\vec{\varrho}_{n-1})\, .
\end{align}
To study the fixed-point, one switches to dimensionless variables as in the previous section and defines $y=\varrho_1^{2/3} t$, to obtain the fixed point equation
\begin{equation}
 \Big\{-d_G - \varrho_i \p_{\varrho_i} - \frac{J_{*}(0)}{3}\,|\vec{\varrho}_1|^{4/3} \,|y| \Big\}
 {G}^{(n)}_{\alpha_1\dots\alpha_{n}}({y,\varrho_1,\vec{n}_1}, \cdots, {\varrho_{n-1},\vec{n}_{n-1}}) =0\, ,
\end{equation}
where hat symbols have been dropped.
 This equation can be integrated introducing $u_1 \equiv \ln{\varrho_1}$ and $u_{i > 1} \equiv \ln{\varrho_1} - \ln{\varrho_i}$. 
 One obtains in terms of the original variables  and including the sub-leading contributions 
 corresponding to the Kolmogorov part as previously
 \begin{align}
 \log\Big( \varepsilon^{\frac{\bar{m}-m}{3}}L^{-d_G}{G}^{(n)}_{\alpha_1\dots\alpha_{n}}(t,{\vp_1}, \cdots, {\vp_{n-1}})\Big)&=  
 -\alpha_L\varepsilon^{1/3} L^{4/3} \,  |t|\, \varrho_1^2
 -d_G \log (\varrho_1 L) \nonumber\\
 & + {{F_L}^{(n)}}_{\alpha_1\dots\alpha_{n}}\left(\varrho_1^{2/3}\varepsilon^{1/3} t,\frac{\vec{\varrho}_{1}}{\varrho_1}, \cdots, \frac{\vec{\varrho}_{n-1}}{\varrho_1}\right)
 + {\cal O}(p_{\rm max} L)\, ,
\end{align}
with $\varrho_1 = \sum_k \vp_k$, and  $\alpha_L = \gamma \hat J_*(0)/4$. As in the previous section, the matrix ${\cal R'}$ can be explicitly constructed for each $n$.
Note that in the more general case of $t_i\equiv t$ for $i=1,\cdots, n'$ and $t_i\equiv 0$ for $i=n'+1,\cdots, n-1$, the above procedure would also lead to a solution with ${\cal R}'$ a $n'\times n'$ matrix 
leaving the $p_i$ for $i=n'+1,\cdots, n-1$ invariant.

As an example, let us specialize to the  case of $G^{(3)}$. One can then use  $\mathcal{R}' = \big( \begin{smallmatrix} 1 &-1\\ 1&1 \end{smallmatrix} \big)$ as
 the matrix from the $\vec{\rho}_i$ to the $\vp_i$, so that
\begin{align}
 \log \big(\varepsilon^{-1}L^{-7}{G}^{(3)}_{\alpha\beta\gamma}({t_1,\vp_1}, {t_{2},\vp_{2}}) \big)&=
  -\alpha_{L}\varepsilon^{1/3} L^{4/3}\,  \abs{t} \abs{\vp_1 + \vp_2}^2 
 - 7 \log( \abs{\vp_1 + \vp_2} L)\nonumber\\
 &+ {{F_L}^{(3)}}_{\alpha\beta\gamma}\Bigg( \abs{\vp_1 + \vp_2}^{2/3} \varepsilon^{1/3}t,\frac{\vp_1+ \vp_2}{\abs{\vp_1 + \vp_2}}, 
 \frac{\vp_2 - \vp_1}{\abs{\vp_1 + \vp_2}}\Bigg) + {\cal O}(p_{\rm max} L)\, .
\end{align}
Again, the simple prediction 
\begin{align}
 &\log \big[\varepsilon^{-1} L^{-7}{G}^{(3)}_{\alpha\beta\gamma}({t,\vp_1}, {t,\vp_{2}})\big] = -\alpha_{L}\varepsilon^{1/3} L^{4/3}\,  \abs{t} \abs{\vp_1 + \vp_2}^2  +  {\cal O}(p_{\rm max}L)\, ,
\end{align}
could be tested in direct numerical simulations of \NS equation.

\end{widetext}

\bibliographystyle{prsty}

\end{document}